\documentclass[journal]{IEEEtran}
\usepackage{amsmath,amsfonts}
\newcommand{\Mod}[1]{\ (\mathrm{mod}\ #1)}
\usepackage{algorithm}
\usepackage{booktabs}
\usepackage{algpseudocode}
\usepackage{array}
\usepackage{amssymb}
\usepackage{enumitem}
\usepackage[caption=false,font=normalsize,labelfont=sf,textfont=sf]{subfig}
\usepackage{textcomp}
\usepackage{stfloats}
\usepackage{url}
\usepackage{verbatim}
\usepackage{graphicx}
\usepackage{cite}
\usepackage{tikz}
\usepackage{graphicx}
\usepackage{float}
\usepackage{framed}
\usepackage[hang,flushmargin]{footmisc}
\usepackage{amsthm}
\newtheorem{theorem}{Theorem}

\newtheorem{corollary}[theorem]{Corollary}
\newtheorem{lemma}[theorem]{Lemma}

\newtheorem{remark}[theorem]{Remark}
\newtheorem{example}{Example}
\DeclareMathOperator{\im}{Im}

\DeclareMathOperator{\wt}{wt}
\DeclareMathOperator{\re}{Re}
\def\N{\mathbb{N}}
\def\Z{\mathbb{Z}}

\def\R{\mathbb{R}}

\newcommand{\overbar}[1]{\mkern 1.5mu\overline{\mkern-1.5mu#1\mkern-1.5mu}\mkern 1.5mu}
\allowdisplaybreaks

\begin{document}

\title{On Codes over Eisenstein Integers}

\author{Abdul Hadi, Uha Isnaini, Indah Emilia Wijayanti, and Martianus Frederic Ezerman
\thanks{Abdul Hadi, Uha Isnaini, and Indah Emilia Wijayanti are with the Department of Mathematics, Universitas Gadjah Mada, Sekip Utara BLS 21, Yogyakarta, 55281, Indonesia (e-mail: abdulhadi1989@mail.ugm.ac.id; isnainiuha@ugm.ac.id; ind\_wijayanti@ugm.ac.id).}
\thanks{Abdul Hadi is also with the Department of Mathematics, Universitas Riau, Tampan, Pekanbaru, 28293, Indonesia (e-mail: abdulhadi@lecturer.unri.ac.id).}
\thanks{Martianus Frederic Ezerman is with the School of Physical and Mathematical Sciences, Nanyang Technological University, Singapore 637371, (e-mail: fredezerman@ntu.edu.sg).}

}



\maketitle

\begin{abstract}
We propose constructions of codes over quotient rings of Eisenstein integers equipped with the Euclidean, square Euclidean, and hexagonal distances as a generalization of codes over Eisenstein integer fields. By set partitioning, we effectively divide the ring of Eisenstein integers into equal-sized subsets for distinct encoding. Unlike in Eisenstein integer fields of prime size, where partitioning is not feasible due to structural limitations, we partition the quotient rings into additive subgroups in such a way that the minimum square Euclidean and hexagonal distances of each subgroup are strictly larger than in the original set. This technique facilitates multilevel coding and enhances signal constellation efficiency.
\end{abstract}


\section{Introduction}
\IEEEPARstart{C}{odes} over finite rings of integers have been extensively studied, with notable contributions found in works such as \cite{Blake1972, Blake1975, Spiegel1977, Calderbank1995, Morita2006}. Many authors have investigated codes over various rings other than $\mathbb{Z}_{n}$ for higher efficiency and reliability in communication and storage.

Codes over Gaussian integers equipped with the Mannheim metric and over Eisenstein integers with the hexagonal metric were explored in \cite{Huber1994} and \cite{Huber1994a}, respectively. The original idea was to consider a finite field, whose elements are Gaussian or Eisenstein integers, as a residue field of the corresponding algebraic integer ring modulo a prime. The next step was to find an element with the smallest norm in each residue class by Euclidean division. This element would then be used to represent an element of the finite field. Codes over algebraic integer rings of cyclotomic fields were investigated in \cite{Fan2004}, where a Mannheim weight based on the minimal Manhattan metric is introduced, along with related topics.

Extending the studies on the codes in \cite{Huber1994}, codes over a finite ring of Gaussian integers were discussed in the series of works in \cite{Ghaboussi2010, Freudenberger2013a, Freudenberger2013, Freudenberger2014}. Further extension to codes over Hurwitz integers was the focus in \cite{Guezeltepe2013, Rohweder2021}. Duran proposed an algebraic construction method for codes over Hurwitz integers in \cite{Duran2023}. To the best of our knowledge, there has not been any extension on the construction of codes over Eisenstein integer fields from \cite{Huber1994a} to codes over a finite ring of Eisenstein integers. 

\textit{Set partitioning} is a coding technique that divides a set of codewords into smaller subsets of specific properties that can be exploited for error correction and detection. The aim is to reduce the search space for symbol detection. Another technique, called \textit{multilevel coding}, combines coding and modulation. \textit{Signal constellation}, which is a set of points in a complex plane that represent the symbols, is usually chosen such that the center of mass of the points is at the origin, is split into subsets of equal size, and the numbering of the subsets is encoded by different codes. This improves on the efficiency and reliability of data transmission. Multilevel coding for a finite ring of Gaussian integers were discussed in \cite{Freudenberger2013, Freudenberger2014}, of Lipschitz integers in \cite{Freudenberger2015}, and of Hurwitz integers in \cite{Rohweder2021, Stern2023}. Set partitioning and multilevel coding over Eisenstein integers with ternary codes were discussed in \cite{Stern2019}. The work introduced a mapping from ternary vectors onto the Eisenstein integers using a modulo function with hexagonal boundaries. In \cite{Stern2019a}, binary multilevel coding over Eisenstein integers was considered for MIMO broadcast transmission. Again, to our best knowledge, there has not been any significant research on set partitioning and multilevel coding for a finite ring of Eisenstein integers.

Most of the linear code constructions, \textit{e.g.}, those proposed in \cite{Huber1994a, Gao2011, Thiers2020, Thiers2021, Thiers2021b}, are based on Eisenstein integer fields of size $q$ built from Eisenstein primes of norm $q \equiv 1 \Mod{3}$, with $q$ being a prime integer. Eisenstein integer fields of size $p^2$ were constructed from a prime number $p \equiv 2 \Mod{3}$ in \cite{Sajjad2023c}. Set partitioning is structurally not feasible on Eisenstein integer fields of prime cardinality. It is also not a good idea to apply on Eisenstein integer fields of squared prime cardinality as partitioning does not increase the minimum distance. 

To address these limitations, we construct codes over finite rings of Eisenstein integers and consider codes over Eisenstein integer fields a particular case. On Eisenstein integers, set partitioning helps in the design of efficient coding schemes that leverage on the unique geometric properties of the integers. Here, it is possible to create multilevel codes akin to the standard coded modulation schemes in \cite{Imai1977,Ungerboeck1982}. Inspired by the hexagonal metric for linear codes over Eisenstein integer fields in \cite{Gao2011}, we show  how to study linear codes over a finite ring of Eisenstein integers equipped with a hexagonal metric and compare their performance with codes over a finite ring of Gaussian integers equipped with a Mannheim metric from \cite{Huber1994, Martinez2007}.

We know from \cite{Conway2013} that Eisenstein integers form the densest packing in a two-dimensional lattice, also known as hexagonal lattices. The optimal packing density can be beneficial in minimizing interference and maximizing data throughput. The points in the lattice, however, are very close to each other, reduce the overall error correction capability. The hexagonal lattice structure of Eisenstein integers allows for an efficient packing of signal points, leading to higher spectral efficiency. More data can then be transmitted within the same bandwidth, improving the overall performance. The ring of Eisenstein integers provides the best performance when used over the AWGN network since it is the best packing in $\R^{2}$. The aim of partitioning is to optimize two metrics, namely, the minimum Euclidean distance and the hexagonal distance in the subsets. We confirm that the hexagonal metric is a suitable distance measure for the set partitioning of Eisenstein integers.

In coding and communication contexts, this high density can be challenging because closely packed signals are more susceptible to interference and errors. While Eisenstein integers are very efficient in terms of space usage, they are not necessarily ideal for applications requiring high error correction capabilities. To address this problem, we use set partitioning to increase the distance between any two distinct points.

Many digital modulation schemes use conventional constellation signals that lack algebraic group, ring, or field structures. Constellations possessing nice algebraic structures can enhance system performance. They enable the constructions of error-correcting codes over complex-valued alphabets. Research on two-dimensional signal constellations has been conducted in the context of phase shift keying (PSK) in \cite{Loeliger1991}, over the Gaussian integers in \cite{Huber1994, Freudenberger2013,Freudenberger2014}, and over the Eisenstein integers in \cite{Huber1994a,Tunali2015,Freudenberger2017}. The respective algebraic properties improve the system design through efficient error correction and low-complexity detection or decoding schemes. These have practical implications in wireless communication systems. Notably, many well-known coding techniques for linear codes leverage these algebraic properties \cite{Freudenberger2013,NobregaNeto2001,Martinez2007,Sun2013,Fang2014,Rohweder2018}.

There have been numerous publications on codes over Gaussian integers, Eisenstein integers, and other cyclotomic fields. Prominent examples include the works done in  \cite{Freudenberger2013,Fang2014,Asif2017,Thiers2021,Thiers2021b}. The known constellations, however, have seen limited practical applications. One possible reason for this lack of adoption is that their number of signal points are typically not a power of $2$. More recent research has explored nonbinary hexagonal modulation schemes, which have been proposed for applications in multi carrier modulation \cite{Han2007}, multilevel coded modulation \cite{Tanahashi2009}, wireless video transmission \cite{Yang2014}, generalized spatial modulation \cite{Freudenberger2017}, physical-layer networking coding \cite{Feng2013,Sun2013}, and MIMO transmission \cite{Tunali2015,Stern2016}.  Eisenstein integers may have high potential in constellation design for future communication systems due to their hexagonal constellation. 

In this paper we aim to carefully select finite rings of Eisenstein integers to use as the alphabet sets for linear codes. 

\noindent
\textbf{Our Contributions}:
\begin{itemize}
\item We introduce two representative classes of finite rings of Eisenstein integers. First, in Theorem \ref{classequivringfac} and Algorithm \ref{algcrs}, we provide a method to find a representative class of finite rings of Eisenstein integers whose elements have real part and the coefficient of $\rho$ is nonnegative. From the representative class, we find another representative class with the smallest norm by using some modulo function. We can reverse the direction, going back from the second class to the first one.  This generalizes known treatment on the Eisenstein integer field of order prime $q\equiv 1\Mod{3}$ from \cite{VazquezCastro2014,vazquez2014}, which can be used in the physical layer alphabet network coding.

\item Freudenberger, Ghaboussi, and Shavgulidze in \cite{Freudenberger2013} demonstrated that a signal constellation of size $n$ in one-dimensional signal space can be transformed into Gaussian constellations only if $n$ is not divisible by the product of  $4$ and a prime integers $p\equiv 3\Mod{4}$ .  Hence, a signal constellation of size $n$  for, {\it e.g.}, $n\in \{3, 7, 12, 19, 21, 27, 28 \}$ cannot be transformed to a Gaussian constellation. In fact, the number of signal points in any signal constellation over the Gaussian integers must be the sum of two squares. Over the Eisenstein integers, the number of signal points does not have such a constraint. 

We provide a necessary and sufficient condition for a Eisenstein integer to be primitive in Theorem \ref{jhjpsiku} and Corollary \ref{transfrizntozeis}. We can then construct a finite ring of Eisenstein integers from a ring of integers modulo $n$, when $n$ is the norm of a primitive Eisenstein integer. Hence, elements in $\Z_{n}$ can be transformed to complex numbers via some modulo function if $n$ is not divisible by the product of  9 and a prime $p\equiv 2 \Mod{3}.$ Thus, we can transform a signal constellation of size $n$  for, {\it e.g.}, $n\in \{3, 7, 12, 19, 21, 27, 28 \}$  into Eisenstein constellations, closing the gap in \cite{Freudenberger2013} entirely.  
    
\item We propose a set partitioning over the Eisenstein integers based on additive groups. Inspired by the set partitioning over the Gaussian integers in \cite{Freudenberger2013,Freudenberger2014}, we give a treatment on the partitioning in two cases based on whether a given Eisenstein integer is primitive or not.

\item We determine the properties of the hexagonal weight and its relation to the norm in Theorem \ref{relnormhex}.   
\item We compare Eisenstein and Gaussian constellations of the same cardinality in terms of average energy using various distance metrics in Table \ref{compcodegausEissame}. These include the Euclidean, squared Euclidean, Mannheim, and hexagonal distances.
\end{itemize}

In terms of organization, Section \ref{sec:two} reviews fundamental properties and provides useful new results on the Eisenstein integers and the hexagonal metric. Section \ref{sec:three} discusses codes over a finite ring of Eisenstein integers. In section \ref{sec:fourth}, we discuss signal constellation over Eisenstein integers and explain the concept of set partitioning based on additive subgroups. Comparisons on the Eisenstein and Gaussian constellations of the same cardinality in term of their average energy requirement are provided in this section. Section \ref{sec:conclu} contains concluding remarks. To keep the exposition smooth, the proofs of some theorems and corollaries are given in the appendices instead of in the main body of the paper. 

\section{Ring of Eisenstein integers}\label{sec:two}

This section discusses known properties of Eisenstein integers and presents new results. These include necessary and sufficient conditions for two classes in a finite ring of Eisenstein integers to be equivalent and for an Eisenstein integer be a primitive. We examine the properties of Euclidean and hexagonal distances and review the division algorithm to define alphabet sets and signal constellations.

\subsection{Arithmetic of Eisenstein integers}
An Eisenstein integer is a complex number $a+b\rho$ with $a,b\in\Z$ and $\rho =\frac{-1}{2}+\frac{\sqrt{3}}{2}i$ being a primitive third root of unity. The set $\mathbb{Z}[\rho ]$ of all Eisenstein integers form a commutative ring under addition and multiplication defined by
\begin{align*}
(a+b\rho )+(c+d\rho )&=(a+c)+(b+d)\rho  \\ 
(a+b\rho )\cdot (c+d\rho )&=(ac-bd)+(ad+bc-bd)\rho. 
\end{align*}
The \emph{conjugate} and \emph{norm} of $\alpha=a+b\rho \in \mathbb{Z}[\rho]$ are 
\begin{align*}
\overbar{\alpha }&=(a-b)-b\rho \;\;\mbox{and}\;\;N_{\rho}(\alpha ) ={{a}^{2}}+{{b}^{2}}-ab\in \mathbb{Z}.
\end{align*}
We have $N_{\rho}(\alpha)=\|\alpha\|^2=\alpha\overbar{\alpha}=N_{\rho}(\overbar{\alpha})$. The norm is multiplicative, that is, 
\begin{equation*}
N_{\rho}(\alpha \, \theta) = N_{\rho}(\alpha) \, N_{\rho}(\theta) \mbox{ for all } \alpha,\theta\in \Z[\rho].
\end{equation*}

Division algorithm works in $\mathbb{Z}[\rho]$, {\it i.e.}, for every $\alpha, \eta \in \mathbb{Z}[\rho]$ such that $\eta\neq 0$, there are unique quotient $ \theta$ and remainder $\delta$ in $\mathbb{Z}[\rho]$ satisfying $\alpha=\theta \, \eta+\delta$ with $N_{\rho}(\delta )<N_{\rho}(\eta )$. The ring $\mathbb{Z}[\rho]$ is a Euclidean domain, a principal ideal domain, and a unique factorization domain.

Given $\alpha ,\eta \in \mathbb{Z}[\rho ]$, an element $\eta $ divides $\alpha$, denoted by $\eta \mid\alpha$ if there is a $\theta \in \mathbb{Z}[\rho ]$ such that $\alpha =\theta \eta$. We say that $\alpha$ is a \emph{unit} if $\alpha \theta =1$ for some $\theta \in \mathbb{Z}[\rho ]$, that is, $\alpha $ has a multiplicative inverse in $\mathbb{Z}[\rho ]$. Elements $\alpha$ and $\beta$ are \emph{associates}, denoted by $\alpha \thicksim \beta$, if $ \alpha=\theta \, \beta$ for some unit $\theta \in \mathbb{Z}[\rho]$. There are $6$ units in $ \Z[\rho] $, namely $ \pm1,\pm\rho$, and $\pm \rho^{2}$. The associates of $\alpha$ are $ \pm\alpha,\pm \rho\alpha $ and $ \pm \rho^{2}\alpha$. 

An Eisenstein integer $\alpha=a+ b \, \rho$ is a \emph{primitive} if $\gcd(a,b)=1$. It is a \emph{prime} if $\alpha $ cannot be expressed as $ \alpha=\theta \, \eta $ with neither $\theta$ nor $\eta$ being a unit in $\Z[\rho] $. Otherwise, $ \alpha $ is \emph{composite}. 

\begin{example}
We can quickly verify that $2$, $5$, $1-\rho$, and $2+ 3\rho$ are primes in $ \Z[\rho]$, while $3$ and $8+5\rho$ are not primes in $\Z[ \rho]$ since $3=(2+\rho)(1-\rho)$ and $8+5\rho=(1-2\rho)(2+3\rho)$.   
\end{example}
 
\begin{theorem}[\cite{Ireland1990}, Classification of Eisenstein Primes]\label{clasprime}
Any prime $\gamma$ in $\mathbb{Z}[\rho]$ is an associate of a prime belonging to one of three types.
\begin{enumerate}[leftmargin=3.5em]
\item[Type 1] contains the prime $1-\rho $.
\item[Type 2] contains primes $a+b \, \rho $ with $N_{\rho}(a+b \, \rho )=q$ being a prime integer and $q\equiv 1\Mod{3}$.
\item[Type 3:] contains prime integers $p$ with $p \equiv 2 \Mod{3}$.
\end{enumerate}
\end{theorem}

Henceforth, we fix $\beta:=1-\rho$ and we let $p$ and $q$ be prime integers such that $p \equiv 2\Mod{3}$ and $q=\psi \, \overbar{\psi}\equiv 1\Mod{3}$, where $\psi$ and $\overbar{\psi}$ are nonassociate Eisenstein primes. We denote by $\gamma$ a generic Eisenstein prime and note that Eisenstein primes $\beta$ and $\psi$, up to associates, are primitive Eisenstein integers.

\subsection{On the Quotient Rings of Eisenstein integers}

Since $\Z[\rho]$ is a principal ideal domain, any ideal is of the form $\langle \eta\rangle$ for some $\eta\in \Z[\rho]$. A congruence in $\Z[\rho]$ modulo $\langle\eta\rangle$ can then be defined. For any $\alpha,\theta\in \Z[\rho]$, $\alpha\equiv \theta\Mod{\eta}$ if and only if $\alpha-\theta\in \langle\eta\rangle$. For any $\alpha\in \Z[\rho]$, the \emph{equivalence class} $[\alpha]_{\eta}$ with respect to $\langle \eta\rangle$ is 
\begin{equation*}
    [\alpha]_{\eta}=\{\theta\in \Z[\rho]\,:\, \theta\equiv \alpha\Mod{ \eta}\}.
\end{equation*} 
The set $\{[\alpha]_{\eta}\,:\, \alpha\in \Z[\rho]\}$ forms the  quotient ring $\Z[\rho]/\langle \eta\rangle$. 

Given $\eta\in \Z[\rho]\setminus\{0\}$, there exists $a+b\rho\in\Z[\rho]$ such that $\eta\sim a+b \, \rho$ and  $\Z[\rho]/\langle\eta\rangle\cong \Z[\rho]/\langle a+b\rho\rangle$. Henceforth, without loss of generality, we assume that $\eta=a+b\rho=t(m+n\rho)$, where $t=\gcd(a,b)$, so that $\gcd(m,n)=1$. 

\begin{theorem}[\cite{Ozkan2013}]\label{factoringiso}
If  $\eta\in\Z[\rho]\setminus\{0\}$ with $\eta=a+b\rho=t(m+n\rho),$ where $\gcd(a,b)=t$ and $\gcd(m,n)=1$, then the complete residue system is
\begin{equation*}
\Z[\rho]/\langle\eta\rangle=\{[x+y\rho]_{\eta}\,:\, 0\le x<tN_{\rho}(m+n\rho),\; 0\le y<t\},
\end{equation*}
with $[x+y\rho]_{\eta}=x+y\rho+\langle\eta\rangle$.
\end{theorem}

Let $\Z_{n}[\rho]=\{a+b\rho\,:\, a,b\in \Z_{n}\}$ be the ring of Eisenstein integers modulo $n$. The next two theorems, reproduced from \cite{Ozkan2013}, establish useful isomorphisms.
 \begin{theorem}[\cite{Ozkan2013}]\label{isomenkripe}
If $ n\geq 1 $ is in $\Z$, then $\Z[\rho]/\langle n\rangle\cong \Z_{n}[\rho]$.
\end{theorem}
\begin{theorem}[\cite{Ozkan2013}]\label{isomeisbul}
If $\eta=a+b\rho$ and $ \gcd(a,b)=1$, then $\Z[\rho]/\langle \eta\rangle\cong \Z_{N_{\rho}(\eta)}$.
\end{theorem}

The two theorems imply the existence of a bijection between $[x+y\rho]_{n} \in \Z[\rho]/\langle n\rangle$ and $[x]_{n}+[y]_{n} \, \rho \in \Z_{n}[\rho]$ and between $[x]_{\eta} \in \Z[\rho]/\langle \eta\rangle$ and $[x]_{N_{\rho}(\eta)} \in \Z_{N_{\rho}(\eta)}$, respectively.

We provide criteria for two classes in $\Z[\rho]/\langle \eta\rangle$, for an arbitrary $\eta$, to be equivalent, generalizing \cite[Theorems 3.4, 3.5, 3.6 and 3.7]{Gullerud2020} that cover the cases where $\eta=\gamma^{k}$ for some Eisenstein prime $\gamma$ and a positive integers $k$.

\begin{theorem}\label{classequivringfac}
Let $\eta=a+b\rho=t(m+n\rho)$ be a nonzero element in $\Z[\rho]$, with $t=\gcd(a,b)$ so that $\gcd(m,n)=1$. Given any $x+y\rho$ and $x'+y'\rho$ in 
$\Z[\rho]$, we have $ [x+y\rho]_{\eta}=[x'+y'\rho]_{\eta} $ if and only if $y'\equiv y\Mod{t}$ and 
\begin{equation*}
\begin{cases}
    n(x'-x)\equiv m(y'-y)\Mod{tN_{\rho}(m+n\,\rho)}\;\;\text{or}\;\\
    m(x'-x)\equiv (n-m)(y-y')\Mod{tN_{\rho}(m+n\,\rho)}.
\end{cases}    
\end{equation*}
In particular, let $0\leq x'<tN_{\rho}(m+n\rho)$, $0\leq y'<t$, and either $m$ or $n$ be relatively prime to $t$. Then $[x+y\rho]_{\eta} = [x'+y'\rho]_{\eta} $ if and only if 
\begin{align}
&y'=y\Mod{t} \mbox{ and} \label{ymodt}\\   
&x'=\begin{cases}
 x-mn^{-1}(y-y')\Mod{tN_{\rho}(m+n\rho)} \\
\qquad \mbox{ if }\gcd(n,t)=1,\\ 
 x+(m^{-1}n-1)(y-y')\Mod{tN_{\rho}(m+n\rho)}\\
 \qquad  \mbox{ if }\gcd(m,t)=1.
\end{cases}\label{pieta}
\end{align}
\end{theorem}
\begin{IEEEproof}
Please see Appendix A.
\end{IEEEproof}

Theorem \ref{classequivringfac} is useful in determining a representative of a class of any Eisenstein integer $\alpha$ in $\Z[\rho]/\langle\eta\rangle$ when $ \eta=a+b\rho=t(m+n\rho) $ and either $m$ or $n$ is relatively prime to $t=\gcd(a,b)$. It is then natural to define the homomorphism 
\begin{equation}\label{pietas}
\pi_{\eta}:\Z[\rho]\rightarrow \Z[\rho]/\langle \eta\rangle    
\end{equation}
that sends 
\[
\pi_{\eta}(x+y\rho) \mapsto [x'+y'\rho]_{\eta} = x'+y'\rho+\langle\eta\rangle
\]
whenever \eqref{ymodt} and \eqref{pieta} are satisfied. 

Algorithm \ref{algcrs} calculates $\pi_{\eta}(\alpha)$ for any $\alpha\in\Z[\rho]$. 
\begin{algorithm}
\caption{A representative class}
\label{algcrs}
\textbf{Input:} Eisenstein integers 
\[
\alpha=x+y\rho \mbox{ and } 
\eta=a+b\rho=t(m+n\rho) \neq 0,
\]
with $t=\gcd(a,b)$ and either $\gcd(m,t)=1$ or $\gcd(n,t)=1$. \\
\textbf{Output:} $x'+y'\rho$ with $0\leq x'<tN_{\rho}(m+n\rho),\,0\leq y'<t$.
\begin{enumerate}
\item[1:] $y' \gets y \Mod{t}$
\item[2:] \textbf{if} $\gcd(n,t)=1$ \textbf{then} \[x' \gets x - m \, n^{-1} (y - y') \Mod{tN_{\rho}(m+n\rho)}\]
\item[3:] \textbf{if} $\gcd(m,t)=1$ \textbf{then} \[x' \gets x + (m^{-1} \, n - 1)  (y - y') \Mod{tN_{\rho}(m+n\rho)}\]
\item[4:] \textbf{return} $x' + y'\rho$
\end{enumerate} 
\end{algorithm}

\begin{remark}
For $ \eta=a+b\rho=t(m+n\rho) $ with $m$ and $n$ being not relatively prime to $t=\gcd(a,b)$, we suggest choosing $\theta\in\Z[\rho]$ with $\theta\sim\eta$ such that $\theta=a'+b'\rho=t'(m'+n'\rho)$ and either $m'$ or $n'$ being relatively prime to $t'=\gcd(a',b')$. For example, let $\eta=12+18\rho=6(2+3\rho)$ with $6=\gcd(12,18)$. Obviously, $6$ is relatively prime to neither $2$ nor $3$. We choose $\eta=18+6\rho$ since $18+6\rho\sim 12+18\rho$.
\end{remark} 

\begin{theorem}\label{psikua}
Let $\eta \sim \psi_{1}^{r_{1}}\psi_{2}^{r_{2}}\cdots \psi_{m}^{r_{m}} = a+b\rho$. Let $N_{\rho}(\psi_{i})=q_{i}\equiv 1\Mod{3}$ be a prime integer for each $i$. If $q_{i} \neq q_{j}$ for $i \neq j$ and $r_{i}\geq 1$, then 
\[
\gcd(a,b)=1 \mbox{ and } \Z[\rho]/\langle \eta\rangle\cong \Z_{N_{\rho}(\eta)}.
\]
\end{theorem}
\begin{IEEEproof}
Appendix B provides a proof.
\end{IEEEproof}
 
\begin{theorem}\label{betapsikua}
Let $\eta\sim\beta\psi_{1}^{r_{1}}\psi_{2}^{r_{2}}\cdots \psi_{m}^{r_{m}}=a+b\rho$, with $\beta :=1-\rho$. Let $N_{\rho}(\psi_{i})=q_{i}\equiv 1\Mod{3}$ be a prime integer for each $i$. If $q_{i}\neq q_{j}$ for $i\neq j$ and $r_{i}\geq 1$, then 
\[
\gcd(a,b)=1 \mbox{ and } 
\Z[\rho]/\langle \eta\rangle\cong \Z_{N_{\rho}(\eta)}.
\]
\end{theorem}

\begin{IEEEproof}
The proof is supplied in Appendix C
\end{IEEEproof}
 
\begin{theorem}\label{jhjpsiku}
An Eisenstein integer $\eta=a+b\rho$ is primitive if and only if the following conditions are met.
\begin{itemize}
    \item $\eta\sim\beta^{r}\psi_{1}^{r_{1}}\cdots \psi_{m}^{r_{m}}$, with $r\in\{0,1\}$ and, for each $i$, $N_{\rho}(\psi_i)=q_i \equiv 1 \Mod{3}$ is a prime,
    \item $q_i\neq q_j$ for $i\neq j$, and
    \item $r_{i}$ and $m$ are nonnegative integers.
\end{itemize}
\end{theorem}
\begin{IEEEproof}
The proof can be found in Appendix D
\end{IEEEproof}

\begin{corollary}\label{transfrizntozeis}
The norm of any primitive Eisenstein integer is not divisible by $9$ and any prime congruent to $2$ modulo $3$.
\end{corollary}
 
\subsection{Distance}
The Euclidean and squared Euclidean distances of Eisenstein integers $\alpha \neq \theta$ are defined, respectively, as 
\begin{align*}
d_{\rm E}(\alpha,\theta) &=\|\alpha-\theta\| \mbox{ and}\\
d_{\rm E}^{2}(\alpha,\theta) &=\|\alpha-\theta\|^{2}=N_{\rho}(\alpha-\theta).
\end{align*}
We can express $\alpha=a+b\rho$ as $\alpha=a_{1}v_{1}+a_{2}v_{2}$ with $v_{1},v_{2}\in \{\pm1,\pm \rho, \pm(1+\rho)\}$. The \emph{hexagonal weight} of $\alpha$ is  
\begin{equation*}
    \wt_{\rm Hex}(\alpha)=\min_{\{a_{1},a_{2}\,:\, a_{1}v_{1}+a_{2}v_{2}=\alpha\}}|a_{1}|+|a_{2}|.
\end{equation*}
More explicitly,
\begin{align}
& \wt_{\rm Hex}(\alpha) = \min\{|a|+|b|,|a-b|+|a|,|a-b|+|b|\} \notag \\
&\quad =\begin{cases}
     |a|+|b|,&\text{if}\;a\leq 0\leq b\;\text{or}\;b\leq 0\leq a,\\
     |a-b|+|a|,&\text{if}\;0\leq a\leq b\;\text{or}\;b\leq a\leq 0,\\
     |a-b|+|b|,&\text{if}\;0\leq b\leq a\;\text{or}\;a\leq b\leq 0.
 \end{cases}
\end{align} 

This weight is the minimum number of unit steps, in directions which are multiples of $\pi/3$ radians, required to reach a certain Eisenstein integer. The weight of a vector is the sum of the weights of all of its entries.

The \emph{hexagonal distance} of two Eisenstein integers is the weight of their difference, {\it i.e.}, 
\[
d_{\rm Hex}(x,y)= \wt_{\rm Hex}(x-y).
\]
The hexagonal distance of two vectors is the sum of the hexagonal distances of the components. Calculating hexagonal distance is more efficient because it does not require computing square roots in the Euclidean distance or multiplications in squared Euclidean distance. The next result is on the properties of the hexagonal weight and its relation to the norm.

\begin{theorem}\label{relnormhex} 
If $\alpha$ and $\theta\in \Z[\rho]$ and $k\in \Z$, then the following statements hold.
\begin{enumerate}
 \item[i)] $\sqrt{N_{\rho}(\alpha)}\leq \wt_{\rm Hex}(\alpha)\leq N_{\rho}(\alpha)$. 
\item[ii)] $\wt_{\rm Hex}(\alpha)\leq\wt_{\rm Hex}(\alpha\theta)$ for all $\theta\in\Z[\rho] \setminus \{0\}$.
\item[iii)] $\wt_{\rm Hex}(\alpha)=\wt_{\rm Hex}(\overbar{\alpha})$.
 \item[iv)] $\wt_{\rm Hex}(k\alpha)= |k|\wt_{\rm Hex}(\alpha)$.
 \item[v)] $\wt_{\rm Hex}(\alpha)=\wt_{\rm Hex}(\theta)=n$ if and only if $\alpha\sim n+k\rho$ and $\theta\sim n+ \ell\rho$ for some $k, \ell \in \{0,1,\ldots,n-1\}$. 
 \item[vi)] If $N_{\rho}(\alpha)=N_{\rho}(\theta)$, then $\wt_{\rm Hex}(\alpha)=\wt_{\rm Hex}(\theta)$. 
\item[vii)] $\wt_{\rm Hex}(\alpha\theta)\leq\wt_{\rm Hex}(\alpha) \, \wt_{\rm Hex}(\theta)$.
\end{enumerate}
\end{theorem}
\begin{IEEEproof}
See Appendix E.
\end{IEEEproof}

\begin{remark}
The converse of (vi) does not hold. To see this we consider $\wt_{\rm Hex}(4+4\rho)=4=\wt_{\rm Hex}(3+4\rho)$. Notice, however, that $N_{\rho}(4+4\rho)=16>13=N_{\rho}(3+4\rho)$. 
Also, $\wt_{\rm Hex}(8+4\rho)=8>7=\wt_{\rm Hex}(7+7\rho)$ but $N_{\rho}(8+4\rho)=48<49=N_{\rho}(7+7\rho)$.    
\end{remark}

\subsection{Division Algorithm}
Jarvis in \cite{Jarvis2013} provided a method, reproduced here as Algorithm \ref{algo1rem}, to compute a remainder when $\alpha$ is divided by $\eta$.
\begin{algorithm}
\caption{A remainder when $\alpha$ is divided by $\eta$, which specifies a closest vector for the ideal lattice $\langle\eta\rangle$}
\label{algo1rem}
\textbf{Input:} Eisenstein integers $\alpha$ and $\eta \neq 0$. \\
\textbf{Output:} A remainder $\delta$ when $\alpha$ is divided by $\eta$.
\begin{enumerate}
\item[1:] $z\gets\tfrac{\alpha}{\eta}=\re(z)+\im(z)i$.
\item[2:] The nearest Eisenstein integers $\theta_{1}$ and $\theta_{2}$ are 
\begin{align*}
\theta_{1}&\gets\lfloor \re(z)\rceil + \left\lfloor \frac{\im(z)}{\sqrt{3}}\right\rceil \, \sqrt{3}i \mbox{ and}\\
\theta_{2}&\gets\lfloor \re(z-\rho)\rceil +\left\lfloor \frac{\im(z-\rho)}{\sqrt{3}} \right\rceil \, \sqrt{3}i+\rho,
\end{align*}
where $\lfloor\cdot\rceil$ denotes the rounding to the nearest integer. More formally, $\lfloor x\rceil=\lfloor x\rfloor$ if $x-\lfloor x\rfloor\leq 1/2$ and $\lfloor x\rceil=\lceil x\rceil$ otherwise.
\item[3:] $\delta_{1}\gets\alpha-\eta \, \theta_{1}$ and $\delta_{2}\gets\alpha-\eta \, \theta_{2}$.
\item[4:] $\delta \gets \alpha \Mod{\eta}$
      \begin{equation}\label{mueta2}
        = \begin{cases}
          \delta_{1}, &\text{if}\quad\|\delta_{1}\|<\|\delta_{2}\|\;\text{or}\;\|\delta_{1}\|=\|\delta_{2}\|\\&\text{and}\; \re(\theta_{1})<\re(\theta_{2}),\\
           \delta_{2}, &\text{if}\quad\|\delta_{2}\|<\|\delta_{1}\|\;\text{or}\;\|\delta_{1}\|=\|\delta_{2}\|\\&\text{and}\; \re(\theta_{2})<\re(\theta_{1}).
      \end{cases}  
      \end{equation}
\end{enumerate} 
\end{algorithm}

Algorithm \ref{algo1rem} gives us the following facts.
\begin{enumerate}
    \item $\alpha=\theta\eta+\delta$ and $N_{\rho}(\delta)<N_{\rho}(\eta)$.
    \item $\theta \, \eta$ is an element in $\langle \eta\rangle=\{\lambda\eta\,:\,\lambda\in\Z[\rho]\}$ that is closest to $\alpha$.
    \item $\delta$ is a smallest representative of its congruence class modulo $\eta$.
    \item The division algorithm guarantees a unique output $\delta$ for any input in $\alpha+\langle \eta\rangle$.
\end{enumerate}

Since the remainder $\delta$ is a smallest representative of the congruence class modulo $\eta$, it always holds that 
\[
N_{\rho}(\delta)\leq N_{\rho}(\alpha) \mbox{ and } 
\wt_{\rm Hex}(\delta)\leq \wt_{\rm Hex}(\alpha).
\]

Fixing a nonzero $\eta\in \Z[\rho]$, we can define the modulo mapping 
\begin{equation}\label{modfunct}
\mu_{\eta} : \Z[\rho] \to \Z[\rho]/\langle \eta\rangle \mbox{ sending }
\alpha \mapsto \delta+\langle \eta\rangle
\end{equation}
for any $\alpha\in\Z[\rho]$, with $\delta=\alpha \Mod{\eta}$ found by Algorithm \ref{algo1rem}. The mapping $\mu_{\eta}$ in \eqref{modfunct} performs a reduction from a particular point in two-dimensional space to its congruent point in the \emph{Voronoi cell} of the Eisenstein integers with respect to the origin and scaled by $\eta$. Given a set of points in a plane, the Voronoi cell for a chosen point is the region consisting of all points which are closer to that point than to any other point in the set.  

\section{Code over a finite ring of Eisenstein integers}\label{sec:three}

This section discusses finite rings of Eisenstein integers as alphabet sets and constructs linear codes over them.

\subsection{Finite Rings as Code Alphabet Sets}

Our alphabet sets are specially crafted finite rings of Eisenstein integers. 

Let $\eta=a+b\rho=t(m+n\rho)$, with $\gcd(a,b)=t$ and $\gcd(m,n)=1$. By Theorem \ref{factoringiso}, we have a complete residue system 
\begin{multline*}
\Z[\rho]/\langle \eta\rangle= \\
\{[x+y\rho]_{\eta} \, : \, 0\leq x<tN_{\rho}(m+n\rho),\;0\leq y<t\},
\end{multline*}
with $[x+y\rho]_{\eta}=x+y\rho+\langle\eta\rangle$. 

Since the kernel of the ring homomorphism $\mu_{\eta}$ in \eqref{modfunct} is $\ker(\mu_{\eta})=\langle \eta\rangle$, we can define the homomorphism 
\[
\pi_{\eta}:\Z[\rho] \to \Z[\rho]/\langle\eta\rangle,
\]
which ensures $\Z[\rho]/\langle \eta\rangle\cong \im(\mu_{\eta})$ under the mapping 
\[
\overbar{\mu}_{\eta}:\Z[\rho]/\langle \eta\rangle \to \Z[\rho]\mbox{, with} 
\]
\begin{equation*}
 \overbar{\mu}_{\eta}([\alpha])=\mu_{\eta}(\alpha)=\alpha \Mod{\eta}\;\;\text{for}\;\;[\alpha]\in \Z[\rho]/\langle \eta\rangle.
\end{equation*} 
Hence, we have a bijection between elements $[x+y\rho]_{\eta}$ in  $\Z[\rho]/\langle \eta\rangle$ and elements $x+y\rho\Mod{\eta}$ in $\im(\mu_{\eta})$. Let \begin{equation*}
\mathcal{R}_{\eta}=\{x+y\rho\,:\,0\leq x<tN_{\rho}(m+n\rho),\;0\leq y<t\}  
\end{equation*}
be a representative point set of the class ring of Eisenstein integers modulo $\eta$. Considering $\mathcal{R}_{\eta}$ as a ring under the modulo addition and multiplication defined by $\pi_{\eta}$, we can denote $\im(\overbar{\mu}_{\eta})$ by \begin{equation*}\label{codealphab}
\mathcal{E}_{\eta}=\{\mu_{\eta}(x+y\rho)\,:\, x+y\rho\in\mathcal{R}_{\eta}\}
\end{equation*}
to represent the factor ring $\Z[\rho]/\langle \eta\rangle$. The ring $\mathcal{E}_{\eta}$ is also a complete residue system modulo $\eta$ equipped with the modulo addition and multiplication defined by $\mu_{\eta}$. The following assertions hold.
\begin{itemize}
\item If $\eta=a+b\rho\sim n\in \N$, then $\gcd(a,b)=n$ and $N_{\rho}(c+d\rho)=1$. Thus,
\begin{equation*}
\mathcal{E}_{\eta}=\{x+y\rho\Mod{\eta}\,:\, 0\leq x<n, \;0\leq y<n \}.
\end{equation*} 
\item If $\eta=a+b\rho\not\sim n\in \N$ and $\gcd(a,b)=1$, then
\begin{equation*}
\mathcal{E}_{\eta}=\{x\Mod{\eta}\,:\, 0\leq x<N_{\rho}(\eta) \}.
\end{equation*}
\item If $\eta=a+b\rho\not\sim n\in \N$ and 
 $\gcd(a,b)=t$, with $1<t<n$, then
\begin{multline*}
\mathcal{E}_{\eta} = \left\{x+y\rho\Mod{\eta} \, : \right.\\
\left. 0\leq x<tN_{\rho}(c+d\rho),\;0\leq y<t \right\}.
\end{multline*}
\end{itemize}

Thus, for every $\alpha\in \mathcal{R}_{\eta}$, we have $ \mu_{\eta}(\alpha)\in \mathcal{E}_{\eta}$. In other words, all elements in $\mathcal{E}_{\eta}$ are images of all elements in $\mathcal{R}_{\eta}$ under $\mu_{\eta}$. Conversely, for a given $\delta\in \mathcal{E}_{\eta}$, we can find $\alpha\in \mathcal{R}_{\eta}$ such that $\mu_{\eta}(\alpha)=\delta$ by using the $\pi_{\eta}$ in \eqref{pietas}. The diagram in \eqref{diagalp} illustrates the transformation of elements in $\mathcal{R}_{\eta}$ to those in $\mathcal{E}_{\eta}$ and vice versa.
\begin{align}
& \mathcal{R}_{\eta} \xrightarrow{\mu_{\eta}}\mathcal{E}_{\eta} \xrightarrow{\pi_{\eta}}\mathcal{R}_{\eta} \mbox{ sending} \notag\\
&x+y\rho \mapsto \mu_{\eta}(x+y\rho) \mapsto \pi_{\eta}(\mu_{\eta}(x+y\rho))=x+y\rho.\label{diagalp}
\end{align}

\begin{example}\label{ex:3}
Let $\eta=-6+5\rho$ and $\alpha=10$. To determine $\alpha\Mod{\eta}$, we calculate the complex numbers
\begin{multline*}
z = \frac{\alpha}{\eta} = \frac{10}{-6+5\rho} 
= \frac{-110}{91} + \frac{-50}{91}\rho 
=\frac{-170}{182}+\frac{-50}{182}\sqrt{3}i\\
\mbox{and } z-\rho =\left(\frac{-170}{182} + \frac{-50}{182}\sqrt{3}i\right)-\frac{-1+\sqrt{3}}{2}i \\
=\frac{-79}{182}+\frac{-141}{182}\sqrt{3}i.
\end{multline*}
Next, we find in $\Z[\rho]$ the respective nearest integers 
\begin{align*}
\theta_{1}&= \left\lfloor \frac{-170}{182}\right\rceil +\left\lfloor \frac{-50}{182} \right \rceil\sqrt{3}i=-1+0\sqrt{3}i=-1 \mbox{ and}\\
\theta_{2}&= \left\lfloor \frac{-79}{182} \right\rceil +\left\lfloor \frac{-141}{182} \right\rceil \sqrt{3}i +\rho \\&= 0-\sqrt{3}i+\rho=-1-\rho.
\end{align*}
We proceed to derive 
\[
\delta_{1}=\alpha-\eta \, \theta_{1}=4+5\rho \mbox{ and } 
\delta_{2}=\alpha-\eta \, \theta_{2}=-1-6\rho.
\]
Since $N_{\rho}(\delta_{1})=21<31=N_{\rho}(\delta_{2})$, in modulo $\eta$ we have
\[
\delta \equiv \alpha =\delta_{1}=4+5\rho.
\]
Thus, $\mu_{\eta}(10) = 10 \Mod{(-6+5\rho)} = 4+5\rho$. 

Conversely, since $\eta=-6+5\rho$, $t=1$, $m=-6$, and $n=5$, we get $y'=5\Mod{1}=0$ and, in modulo $91$, 
\[
x' \equiv 4-(-6)(5)^{-1}(5-0) \equiv 4+6 = 10.
\]
Thus, $\pi_{\eta}(4+5\rho)=10+\langle \eta\rangle$. 
\end{example}

\begin{example}\label{ex:4}
Let $\eta=4+6\rho$ and $\alpha=10+\rho$. We begin by calculating 
\[
z=\frac{\alpha}{\eta} = 
\frac{10+\rho}{4+6\rho}= 
\frac{-14}{28}+ \frac{-56}{28}\rho 
=\frac{1}{2}-\sqrt{3}i.
\]
Hence, 
\[
z-\rho = \frac{1}{2}-\sqrt{3}i - \frac{-1+\sqrt{3}}{2}i = 1-\frac{3}{2}\sqrt{3}i.
\]
The nearest integers $\theta_{1}$ and $\theta_{2}$ in $\Z[\rho]$ are
\begin{align*}
\theta_{1}&= \left\lfloor \frac{1}{2} \right\rceil + \lfloor -1\rceil\sqrt{3}i=-\sqrt{3}i=-1-2\rho \mbox{ and}\\
\theta_{2}&=\lfloor 1\rceil+ \left\lfloor \frac{-3}{2} \right\rceil \sqrt{3}i+\rho = 1-2\sqrt{3}i+\rho=-1-3\rho.
\end{align*}
Hence, $\delta_{1}=\alpha-\eta \, \theta_{1}=2+3\rho$ and $\delta_{2}=\alpha-\eta \, \theta_{2}=-4+\rho$. Since $N_{\rho}(\delta_{1})=7<21=N_{\rho}(\delta_{2})$, we get 
\[
\delta=\alpha\Mod{\eta}=\delta_{1}=2+3\rho.
\]
Thus, $\mu_{\eta}(10+\rho)=10+\rho\Mod{4+6\rho}=2+3\rho$. 

Conversely, since $\eta=4+6\rho=2(2+3\rho)$, we obtain $t=2$, $m=2$, and $n=3$, leading to $y'=3\Mod{2}=1$ and 
\begin{align*}
x'&=2-(2)(3)^{-1}(3-1)\Mod{14}\\
&=2-(2)(5)(2)\Mod{14}\\
&=-4 \Mod{14}=10. 
\end{align*} 
Thus, $\pi_{\eta}(-2-3\rho)=10+\rho+\langle \eta\rangle$. 
\end{example}

\begin{example}\label{ex:5}
Letting $n=6$, $\psi_{1}=2+3\rho$, $\psi_{2}=3+4\rho$, and $\psi_{1}\psi_{2}=-6+5\rho$ such that $N_{\rho}(\psi_{1}\psi_{2})=91$, we get 
\begin{align*}
\mathcal{E}_{6} &\cong \Z_{6}[\rho] \cong \Z[\rho]/\langle 6\rangle,\\
\mathcal{E}_{6+12\rho} &\cong \Z[\rho]/\langle 6+12\rho\rangle {, and}\\
\mathcal{E}_{\psi_{1}\psi_{2}} &\cong \Z_{91}\cong\Z[\rho]/\langle \psi_{1}\psi_{2}\rangle.
\end{align*}
\end{example}

The respective rings from Examples \ref{ex:3} to \ref{ex:5} are presented in Tables \ref{residu6a} to \ref{residupsi1psi2a}.

\begin{table}[!ht]
\caption{Elements in $\Z_{6}[\rho],\;\Z[\rho]/\langle 6\rangle$ and $\mathcal{E}_{6}$}\label{residu6a}
\renewcommand{\arraystretch}{1.1}
\centering
\begin{tabular}{lll}
\toprule
$\Z_{6}[\rho]$& $\Z[\rho]/\langle 6\rangle$  & $\mathcal{E}_{6}$ \\
\midrule
$[0]_{6}$  &  $[0]_{6}$  & $0$ \\
$[1]_{6}$  & $[1]_{6}$  & $1$ \\
         $[2]_{6}$  & $[2]_{6}$  & $2$ \\
          $[3]_{6}$  & $[3]_{6}$  & $3$\\
         $[4]_{6}$  & $[4]_{6}$  & $-2$\\
         $[5]_{6}$  & $[5]_{6}$  & $-1$ \\
         $[1]_{6}\rho$  & $[\rho]_{6}$  & $\rho$ \\
         $[1]_{6}+[1]_{6}\rho$& $[1+\rho]_{6}$& $1+\rho$\\
        $[2]_{6}+[1]_{6}\rho$& $[2+\rho]_{6}$& $2+\rho$\\
      $[3]_{6}+[1]_{6}\rho$& $[3+\rho]_{6}$& $3+\rho$\\
     $[4]_{6}+[1]_{6}\rho$& $[4+\rho]_{6}$& $-2+\rho$\\
     $[5]_{6}+[1]_{6}\rho$& $[5+\rho]_{6}$& $-1+\rho$\\
         $[2]_{6}\rho$  & $[2\rho]_{6}$  & $2\rho$\\
         $[1]_{6}+[2]_{6}\rho$& $[1+2\rho]_{6}$& $1+2\rho$\\
        $[2]_{6}+[2]_{6}\rho$& $[2+2\rho]_{6}$& $2+2\rho$\\
      $[3]_{6}+[2]_{6}\rho$& $[3+2\rho]_{6}$& $3+2\rho$\\
     $[4]_{6}+[2]_{6}\rho$& $[4+2\rho]_{6}$& $4+2\rho$\\
     $[5]_{6}+[2]_{6}\rho$& $[5+2\rho]_{6}$& $-1+2\rho$\\ 
     $[3]_{6}\rho$& $[3\rho]_{6}$& $-3\rho$\\
   $[1]_{6}+[3]_{6}\rho$& $[1+3\rho]_{6}$& $1+3 \rho$\\
 $[2]_{6}+[3]_{6}\rho$& $[2+3\rho]_{6}$& $2+3\rho$\\
  $[3]_{6}+[3]_{6}\rho$& $[3+3\rho]_{6}$&$3+3\rho$\\
  $[4]_{6}+[3]_{6}\rho$& $[4+3\rho]_{6}$& $-2-3\rho$\\
   $[5]_{6}+[3]_{6}\rho$& $[5+3\rho]_{6}$& $-1-3\rho$\\
 $[4]_{6}\rho$& $[4\rho]_{6}$& $-2\rho$\\
$[1]_{6}+[4]_{6}\rho$& $[1+4\rho]_{6}$& $1-2\rho$\\
$[2]_{6}+[4]_{6}\rho$& $[2+4\rho]_{6}$& $2-2\rho$\\
 $[3]_{6}+[4]_{6}\rho$& $[3+4\rho]_{6}$& $-3-2\rho$\\
 $[4]_{6}+[4]_{6}\rho$& $[4+4\rho]_{6}$& $-2-2\rho$\\
   $[5]_{6}+[4]_{6}\rho$& $[5+4\rho]_{6}$& $-1-2\rho$\\
 $[5]_{6}\rho$& $[5\rho]_{6}$& $-\rho$\\
        $[1]_{6}+[5]_{6}\rho$& $[1+5\rho]_{6}$& $1-\rho$\\
       $[2]_{6}+[5]_{6}\rho$& $[2+5\rho]_{6}$& $2-\rho$\\
     $[3]_{6}+[5]_{6}\rho$& $[3+5\rho]_{6}$& $-3-\rho$\\
    $[4]_{6}+[5]_{6}\rho$& $[4+5\rho]_{6}$& $-2-\rho$\\
     $[5]_{6}+[5]_{6}\rho$& $[5+5\rho]_{6}$& $-1-\rho$\\ 
\bottomrule
\end{tabular}
\end{table}

\begin{table}[!ht]
\caption{Elements in $\Z[\rho]/\langle 6+12\rho\rangle $ and $\mathcal{E}_{6+12\rho}$}\label{residu6psibar} 
\renewcommand{\arraystretch}{1.1}
\centering
\begin{tabular}{ll|ll}
\toprule
 $\Z[\rho]/\langle 6+12\rho\rangle $ & $\mathcal{E}_{6+12\rho}$ &$\Z[\rho]/\langle 6+12\rho\rangle $ & $\mathcal{E}_{6+12\rho}$  \\
\midrule
$[0]_{6+12\rho}$ & $0$ &$[3\rho]_{6+12\rho}$  & $3\rho$\\
$[1]_{6+12\rho}$ & $1$ &$[1+3\rho]_{6+12\rho}$ & $1+3\rho$\\
$[2]_{6+12\rho}$ & $2$ &$[2+3\rho]_{6+12\rho}$  & $2+3\rho$\\
$[3]_{6+12\rho}$ & $3$ &$[3+3\rho]_{6+12\rho}$ & $3+3\rho$\\
$[4]_{6+12\rho}$ & $4$ &$[4+3\rho]_{6+12\rho}$  & $4+3\rho$\\
$[5]_{6+12\rho}$ & $5$ &$[5+3\rho]_{6+12\rho}$ & $5+3\rho$\\
$[6]_{6+12\rho}$ & $6\rho $ &$[6+3\rho]_{6+12\rho}$  & $6+3\rho$\\
$[7]_{6+12\rho}$ & $1+6\rho $ &$[7+3\rho]_{6+12\rho}$ & $-5-3\rho$\\
$[8]_{6+12\rho}$ & $2+6\rho $ &$[8+3\rho]_{6+12\rho}$  & $-4-3\rho$\\
$[9]_{6+12\rho}$ & $3+6\rho $ &$[9+3\rho]_{6+12\rho}$ & $-3-3\rho$\\
$[10]_{6+12\rho}$ & $4+6\rho$&$[10+3\rho]_{6+12\rho}$  & $-2-3\rho$\\
$[11]_{6+12\rho}$ & $5+6\rho $&$[11+3\rho]_{6+12\rho}$ & $-1-3\rho$\\
$[12]_{6+12\rho}$ & $6+6\rho $&$[12+3\rho]_{6+12\rho}$  & $-3\rho$\\
$[13]_{6+12\rho}$ & $-5$&$[13+3\rho]_{6+12\rho}$ & $1-3\rho$\\
$[14]_{6+12\rho}$ & $-4$&$[14+3\rho]_{6+12\rho}$& $2-3\rho$\\
$[15]_{6+12\rho}$ & $-3$&$[15+3\rho]_{6+12\rho}$ & $-3+3\rho$\\
$[16]_{6+12\rho}$ & $-2$&$[16+3\rho]_{6+12\rho}$  & $-2+3\rho$\\
$[17]_{6+12\rho}$ & $-1$&$[17+3\rho]_{6+12\rho}$ & $-1+3\rho$\\
$[\rho]_{6+12\rho}$ & $\rho$&$[4\rho]_{6+12\rho}$  & $4\rho$\\
$[1+\rho]_{6+12\rho}$ & $1+\rho$&$[1+4\rho]_{6+12\rho}$ & $1+4\rho$\\
$[2+\rho]_{6+12\rho}$ & $2+\rho$&$[2+4\rho]_{6+12\rho}$  & $2+4\rho$\\
$[3+\rho]_{6+12\rho}$ & $3+\rho$&$[3+4\rho]_{6+12\rho}$ & $3+4\rho$\\
$[4+\rho]_{6+12\rho}$ & $4+\rho$&$[4+4\rho]_{6+12\rho}$  & $4+4\rho$\\
$[5+\rho]_{6+12\rho}$ & $5+\rho$&$[5+4\rho]_{6+12\rho}$ & $5+4\rho$\\
$[6+\rho]_{6+12\rho}$ & $6+\rho$&$[6+4\rho]_{6+12\rho}$  & $6+4\rho$\\
$[7+\rho]_{6+12\rho}$ & $-5-5\rho$&$[7+4\rho]_{6+12\rho}$ & $-5-2\rho$\\
$[8+\rho]_{6+12\rho}$ & $-4-5\rho$&$[8+4\rho]_{6+12\rho}$  & $-4-2\rho$\\
$[9+\rho]_{6+12\rho}$ & $-3-5\rho$&$[9+4\rho]_{6+12\rho}$ & $-3-2\rho$\\
$[10+\rho]_{6+12\rho}$ & $-2-5\rho$&$[10+4\rho]_{6+12\rho}$  & $-2-2\rho$\\
$[11+\rho]_{6+12\rho}$ & $-1-5\rho$&$[11+4\rho]_{6+12\rho}$ & $-1-2\rho$\\
$[12+\rho]_{6+12\rho}$ & $-5\rho$&$[12+4\rho]_{6+12\rho}$  & $-2\rho$\\
$[13+\rho]_{6+12\rho}$ & $-5+\rho$&$[13+4\rho]_{6+12\rho}$ & $1-2\rho$\\
$[14+\rho]_{6+12\rho}$ & $-4+\rho$&$[14+4\rho]_{6+12\rho}$  & $2-2\rho$\\
$[15+\rho]_{6+12\rho}$ & $-3+\rho$&$[15+4\rho]_{6+12\rho}$ & $3-2\rho$\\
$[16+\rho]_{6+12\rho}$ & $-2+\rho$&$[16+4\rho]_{6+12\rho}$  & $-2+4\rho$\\
$[17+\rho]_{6+12\rho}$ & $-1+\rho$&$[17+4\rho]_{6+12\rho}$ & $-1+4\rho$\\
$[2\rho]_{6+12\rho}$ & $2\rho$&$[5\rho]_{6+12\rho}$  & $5\rho$\\
$[1+2\rho]_{6+12\rho}$ & $1+2\rho$&$[1+5\rho]_{6+12\rho}$ & $1+5\rho$\\
$[2+2\rho]_{6+12\rho}$ & $2+2\rho$&$[2+5\rho]_{6+12\rho}$  & $2+5\rho$\\
$[3+2\rho]_{6+12\rho}$ & $3+2\rho$&$[3+5\rho]_{6+12\rho}$ & $3+5\rho$\\
$[4+2\rho]_{6+12\rho}$ & $4+2\rho$&$[4+5\rho]_{6+12\rho}$  & $4+5\rho$\\
$[5+2\rho]_{6+12\rho}$ & $5+2\rho$&$[5+5\rho]_{6+12\rho}$ & $5+5\rho$\\
$[6+2\rho]_{6+12\rho}$ & $6+2\rho$&$[6+5\rho]_{6+12\rho}$  & $6+5\rho$\\
$[7+2\rho]_{6+12\rho}$ & $-5-4\rho$&$[7+5\rho]_{6+12\rho}$ & $-5-\rho$\\
$[8+2\rho]_{6+12\rho}$ & $-4-4\rho$&$[8+5\rho]_{6+12\rho}$  & $-4-\rho$\\
$[9+2\rho]_{6+12\rho}$ & $-3-4\rho$&$[9+5\rho]_{6+12\rho}$ & $-3-\rho$\\
$[10+2\rho]_{6+12\rho}$ & $-2-4\rho$&$[10+5\rho]_{6+12\rho}$  & $-2-\rho$\\
$[11+2\rho]_{6+12\rho}$ & $-1-4\rho$&$[11+5\rho]_{6+12\rho}$ & $-1-\rho$\\
$[12+2\rho]_{6+12\rho}$ & $-4\rho$&$[12+5\rho]_{6+12\rho}$  & $-\rho$\\
$[13+2\rho]_{6+12\rho}$ & $1-4\rho$&$[13+5\rho]_{6+12\rho}$ & $1-\rho$\\
$[14+2\rho]_{6+12\rho}$ & $-4+2\rho$&$[14+5\rho]_{6+12\rho}$  & $2-\rho$\\
$[15+2\rho]_{6+12\rho}$ & $-3+2\rho$&$[15+5\rho]_{6+12\rho}$ & $3-\rho$\\
$[16+2\rho]_{6+12\rho}$ & $-2+2\rho$&$[16+5\rho]_{6+12\rho}$  & $4-\rho$\\
$[17+2\rho]_{6+12\rho}$ & $-1+2\rho$&$[17+5\rho]_{6+12\rho}$ & $-1+5\rho$\\
\bottomrule
\end{tabular}
\end{table}

\begin{table*}[!t]
\caption{Elements in $\Z_{91},\;\Z[\rho]/\langle\psi_{1}\psi_{2}\rangle $ and $\mathcal{E}_{\psi_{1}\psi_{2}}$}
\label{residupsi1psi2a}
\renewcommand{\arraystretch}{1.1}
\centering
\begin{tabular}{ccc|ccc|ccc}
\toprule
 $\Z_{91}$ & $\Z[\rho]/\langle\psi_{1}\psi_{2}\rangle$ & $\mathcal{E}_{\psi_{1}\psi_{2}}$ & $\Z_{91}$ & $\Z[\rho]/\langle\psi_{1}\psi_{2}\rangle$ & $\mathcal{E}_{\psi_{1}\psi_{2}}$ & $\Z_{91}$ & $\Z[\rho]/\langle\psi_{1}\psi_{2}\rangle$ & $\mathcal{E}_{\psi_{1}\psi_{2}}$  \\
 \midrule
        $[0]_{91}$ & $[0]_{\psi_{1}\psi_{2}}$ & $0$ &$[31]_{91}$&$[31]_{\psi_{1}\psi_{2}}$  &$-3-2\rho$  & $[62]_{91}$ & $[62]_{\psi_{1}\psi_{2}}$& $5+2\rho$ \\
         $[1]_{91}$&$[1]_{\psi_{1}\psi_{2}}$ & $1$ &$[32]_{91}$&$[32]_{\psi_{1}\psi_{2}}$ & $-2-2\rho$ & $[63]_{91}$& $[63]_{\psi_{1}\psi_{2}}$ &$-5-4\rho$\\
        $[2]_{91}$ & $[2]_{\psi_{1}\psi_{2}}$ & $2$ & $[33]_{91}$& $[33]_{\psi_{1}\psi_{2}}$ &$-1-2\rho$  & $[64]_{91}$& $[64]_{\psi_{1}\psi_{2}}$ & $-4-4\rho$ \\
        $[3]_{91}$ & $[3]_{\psi_{1}\psi_{2}}$ & $3$ & $[34]_{91}$& $[34]_{\psi_{1}\psi_{2}}$ &$-2\rho$  & $[65]_{91}$& $[65]_{\psi_{1}\psi_{2}}$ &$-3-4\rho$ \\
        $[4]_{91}$ & $[4]_{\psi_{1}\psi_{2}}$ & $4$ & $[35]_{91}$& $[35]_{\psi_{1}\psi_{2}}$ & $1-2\rho$ & $[66]_{91}$ & $[66]_{\psi_{1}\psi_{2}}$ & $-2-4\rho$ \\
        $[5]_{91}$ & $[5]_{\psi_{1}\psi_{2}}$ & $5$& $[36]_{91}$ & $[36]_{\psi_{1}\psi_{2}}$ &$2-2\rho$  & $[67]_{91}$& $[67]_{\psi_{1}\psi_{2}}$ &$-1-4\rho$  \\
        $[6]_{91}$ & $[6]_{\psi_{1}\psi_{2}}$ &$5\rho$  & $[37]_{91}$ & $[37]_{\psi_{1}\psi_{2}}$ & $3-2\rho$ & $[68]_{91}$& $[68]_{\psi_{1}\psi_{2}}$ & $-4\rho$ \\
        $[7]_{91}$ & $[7]_{\psi_{1}\psi_{2}}$ &$1+5\rho$  & $[38]_{91}$& $[38]_{\psi_{1}\psi_{2}}$ & $-2+3\rho$ & $[69]_{91}$& $[69]_{\psi_{1}\psi_{2}}$ & $1-4\rho$ \\
       $[8]_{91}$ &   $[8]_{\psi_{1}\psi_{2}}$ & $2+5\rho$&$[39]_{91}$ &$[39]_{\psi_{1}\psi_{2}}$  & $-1+3\rho$ & $[70]_{91}$& $[70]_{\psi_{1}\psi_{2}}$ & $-4+\rho$ \\
       $[9]_{91}$ &  $[9]_{\psi_{1}\psi_{2}}$ & $3+5\rho$ &$[40]_{91}$&$[40]_{\psi_{1}\psi_{2}}$  & $3\rho$ & $[71]_{91}$& $[71]_{\psi_{1}\psi_{2}}$ & $-3+\rho$ \\
        $[10]_{91}$ & $[10]_{\psi_{1}\psi_{2}}$ &  $4+5\rho$& $[41]_{91}$ & $[41]_{\psi_{1}\psi_{2}}$ & $1+3\rho$ & $[72]_{91}$& $[72]_{\psi_{1}\psi_{2}}$ & $-2+\rho$ \\
        $[11]_{91}$ & $[11]_{\psi_{1}\psi_{2}}$ &  $5+5\rho$& $[42]_{91}$ & $[42]_{\psi_{1}\psi_{2}}$ & $2+3\rho$  & $[73]_{91}$& $[73]_{\psi_{1}\psi_{2}}$ & $-1+\rho$ \\
        $[12]_{91}$ & $[12]_{\psi_{1}\psi_{2}}$ &  $-5-\rho$ & $[43]_{91}$& $[43]_{\psi_{1}\psi_{2}}$ & $3+3\rho$ & $[74]_{91}$& $[74]_{\psi_{1}\psi_{2}}$ &$\rho$  \\
        $[13]_{91}$ & $[13]_{\psi_{1}\psi_{2}}$ & $-4-\rho$ & $[44]_{91}$& $[44]_{\psi_{1}\psi_{2}}$ & $4+3\rho$ & $[75]_{91}$& $[75]_{\psi_{1}\psi_{2}}$ & $1+\rho$ \\
       $[14]_{91}$ &  $[14]_{\psi_{1}\psi_{2}}$ & $-3-\rho$ & $[45]_{91}$& $[45]_{\psi_{1}\psi_{2}}$ & $5+3\rho$ & $[76]_{91}$& $[76]_{\psi_{1}\psi_{2}}$ & $2+\rho$ \\
       $[15]_{91}$ &  $[15]_{\psi_{1}\psi_{2}}$ & $-2-\rho$ & $[46]_{91}$& $[46]_{\psi_{1}\psi_{2}}$ & $-5-3\rho$ & $[77]_{91}$& $[77]_{\psi_{1}\psi_{2}}$ &$3+\rho$  \\
        $[16]_{91}$ &   $[16]_{\psi_{1}\psi_{2}}$ &$-1-\rho$ & $[47]_{91}$ & $[47]_{\psi_{1}\psi_{2}}$ &$-4-3\rho$  & $[78]_{91}$& $[78]_{\psi_{1}\psi_{2}}$ & $4+\rho$ \\
        $[17]_{91}$ & $[17]_{\psi_{1}\psi_{2}}$ & $-\rho$& $[48]_{91}$ & $[48]_{\psi_{1}\psi_{2}}$ & $-3-3\rho$& $[79]_{91}$ & $[79]_{\psi_{1}\psi_{2}}$ & $5+\rho$ \\
        $[18]_{91}$ & $[18]_{\psi_{1}\psi_{2}}$ & $1-\rho$ & $[49]_{91}$ & $[49]_{\psi_{1}\psi_{2}}$ & $-2-3\rho$ & $[80]_{91}$ & $[80]_{\psi_{1}\psi_{2}}$ & $-5-5\rho$ \\
      $[19]_{91}$ &$[19]_{\psi_{1}\psi_{2}}$ & $2-\rho$& $[50]_{91}$ & $[50]_{\psi_{1}\psi_{2}}$ & $-1-3\rho$ & $[81]_{91}$& $[81]_{\psi_{1}\psi_{2}}$ & $-4-5\rho$ \\
      $[20]_{91}$ &   $[20]_{\psi_{1}\psi_{2}}$ & $3-\rho$& $[51]_{91}$ & $[51]_{\psi_{1}\psi_{2}}$ &$-3\rho$ & $[82]_{91}$ & $[82]_{\psi_{1}\psi_{2}}$ &$-3-5\rho$  \\
     $[21]_{91}$ &    $[21]_{\psi_{1}\psi_{2}}$ & $4-\rho$  & $[52]_{91}$& $[52]_{\psi_{1}\psi_{2}}$ & $1-3\rho$ & $[83]_{91}$& $[83]_{\psi_{1}\psi_{2}}$ & $-2-5\rho$ \\
      $[22]_{91}$ &   $[22]_{\psi_{1}\psi_{2}}$ &$-1+4\rho$  & $[53]_{91}$& $[53]_{\psi_{1}\psi_{2}}$ & $2-3\rho$ & $[84]_{91}$& $[84]_{\psi_{1}\psi_{2}}$ & $-1-5\rho$ \\
      $[23]_{91}$ &    $[23]_{\psi_{1}\psi_{2}}$ & $4\rho$ &$[54]_{91}$&$[54]_{\psi_{1}\psi_{2}}$  & $-3+2\rho$ & $[85]_{91}$& $[85]_{\psi_{1}\psi_{2}}$ & $-5\rho$ \\
      $[24]_{91}$ &   $[24]_{\psi_{1}\psi_{2}}$ & $1+4\rho$ &$[55]_{91}$ & $[55]_{\psi_{1}\psi_{2}}$ & $-2+2\rho$ & $[86]_{91}$& $[86]_{\psi_{1}\psi_{2}}$ &$-5$ \\
      $[25]_{91}$ &   $[25]_{\psi_{1}\psi_{2}}$ & $2+4\rho$ & $[56]_{91}$ &$[56]_{\psi_{1}\psi_{2}}$& $-1+2\rho$ & $[87]_{91}$& $[87]_{\psi_{1}\psi_{2}}$ & $-4$ \\
      $[26]_{91}$ &   $[26]_{\psi_{1}\psi_{2}}$ & $3+4\rho$  & $[57]_{91}$ & $[57]_{\psi_{1}\psi_{2}}$& $2\rho$ & $[88]_{91}$& $[88]_{\psi_{1}\psi_{2}}$ & $-3$ \\
      $[27]_{91}$ &   $[27]_{\psi_{1}\psi_{2}}$ & $4+4\rho$  & $[58]_{91}$& $[58]_{\psi_{1}\psi_{2}}$ & $1+2\rho$ & $[89]_{91}$& $[89]_{\psi_{1}\psi_{2}}$ &$-2$  \\
       $[28]_{91}$ &  $[28]_{\psi_{1}\psi_{2}}$ & $5+4\rho$ & $[59]_{91}$ & $[59]_{\psi_{1}\psi_{2}}$& $2+2\rho$ & $[90]_{91}$& $[90]_{\psi_{1}\psi_{2}}$ & $-1$ \\
       $[29]_{90}$ &  $[29]_{\psi_{1}\psi_{2}}$ & $-5-2\rho$ & $[60]_{90}$ & $[60]_{\psi_{1}\psi_{2}}$& $3+2\rho$  \\
      $[30]_{90}$ &   $[30]_{\psi_{1}\psi_{2}}$ & $-4-2\rho$ & $[61]_{90}$ & $[61]_{\psi_{1}\psi_{2}}$& $4+2\rho$   \\
\bottomrule
\end{tabular}
\end{table*}

\subsection{Fields and Field Extensions of Eisenstein Integers}
For the different primes in $\Z[\rho]$ we have finite fields
\begin{align*}
    \Z[\rho]/\langle \gamma\rangle\cong \begin{cases}
        \Z_{3},\;\;&\;\text{if}\;\gamma\sim \beta\\
        \Z_{q},\;\;&\;\text{if}\;\gamma\sim \psi\;\;\text{with prime}\\ 
        &\;\quad q=N_{\rho}(\psi)\equiv 1\Mod{3}\\
          \Z_{p}[\rho],\;\;&\;\text{if}\;\gamma\sim p\;\;\text{with prime } p\equiv 2\Mod{3}.
    \end{cases}
\end{align*}
For any Eisenstein field $R_{\gamma}$ or $\mathcal{E}_{\gamma}$ for a fixed prime $ \gamma\in\Z[\rho]$, we can construct a Euclidean domain $R_{\gamma}[X]$ and, subsequently, an Eisenstein field extension $ R_{\gamma}^{n} $, which is the quotient ring 
\[ 
R_{\gamma}[X]/\langle f(X)\rangle.
\]
The maximal ideal $ \langle f(X)\rangle $ is generated by an irreducible polynomial $ f(X)$ of degree $n$ in $ R_{\gamma}[X]$. If $\alpha$ denotes the coset $ X+\langle f(X)\rangle$, then $ f(\alpha)=0 $ and the corresponding Eisenstein field extension is
\[
R_{\gamma}^{n}=
\left\{ \sum_{j=0}^{n-1} a_{j} \, \alpha^{j} \, : \, a_{j} \in R_{\gamma}\right\}.
\]

\begin{example}
The polynomial $f(X)=X^{2}+X+\rho $ is primitive in $R_{2}[X]$. If $ \alpha $ is a root of $ f(X) $ over $R_{2}$ then $f(\alpha)=\alpha^{2}+\alpha+\rho=0$. Thus, 
$\alpha^{2}=\alpha+\rho$,
\[ R_{2}^{2}=\{a_{0}+a_{1}\alpha \, : \, a_{0},a_{1}\in \mathcal{R}_{2}\}
\]
and $ (R_{2}^{2})^{\ast}=R_{2}^{2}\setminus\{0\}$ is a multiplicative cyclic group of order $ 2^{4}-1=15$.
\end{example}
Generally, the $n$-degree extension field $R_{\gamma}^{n}$ of $ R_{\gamma}$ is of order $ N_{\rho}(\gamma)^{n}$ and  $(R_{\gamma}^{n})^{\ast}:=R_{\gamma}^{n}\setminus\{0\} $ is a multiplicative cyclic group of order $N_{\rho}(\gamma)^{n}-1$.

\subsection{Linear Codes over a Finite Ring of Eisenstein Integers}

Let $R_{\eta}=\mathcal{R}_{\eta}$ or $\mathcal{E}_{\eta}$ be a set of elements in the quotient ring $\Z[\rho]/\langle \eta\rangle$. We use $R :=R_{\eta}$ with $N_{\rho}(\eta)$ elements as the underlying alphabet set of our codes. A code is a nonempty set $C \subseteq R^{n}$ and its elements are called \emph{codewords}. A linear code $C$ of length $n$ over $R$ is defined to be a submodule of $R^{n}$.

Since $R$ and $R^{n}$ are abelian groups, we say that $C$ is a \emph{group code} if it is a subgroup of $R^{n}$. Equivalently, $c,c'\in C \implies c-c'\in C$. When $R$ is a finite field, {\it i.e.}, $R^{n}$ is a vector space of dimension $n$ over $R$, a linear code $C$ is a subspace of $R^{n}$. We call $C$ an $(n,k)$ code if $C$ has exactly $|R|^{k}$ codewords.

\section{Signal Constellation}\label{sec:fourth}

We label by $\mathcal{E}_{\eta}$ an \emph{Eisenstein constellation} of $N_{\rho}(\eta)$ elements. Our aim is to minimize the total squared Euclidean and hexagonal distances. The \emph{average energy} of a constellation is the expected energy when all elements are used with equal
probability. Hence, the average energy of $\mathcal{E}_{\eta}$ based on the squared Euclidean and the hexagonal distances are given respectively by
\begin{align}
E^{2}(\mathcal{E}_{\eta}) &= \frac{1}{N_{\rho}(\eta)} \sum_{\alpha\in \mathcal{E}_{\eta}}N_{\rho}(\alpha) \mbox{ and}\\
E_{\rm Hex}(\mathcal{E}_{\eta}) &=\frac{1}{N_{\rho}(\eta)}\sum_{\alpha\in \mathcal{E}_{\eta}} \wt_{\rm Hex}(\alpha).
\end{align}
Since, by Theorem \ref{relnormhex} Part i), $\wt_{\rm Hex}(\eta)\leq N_{\rho}(\eta)$, we have 
\[
E_{\rm Hex}(\mathcal{E}_{\eta})\leq  E^{2}(\mathcal{E}_{\eta}).
\]
Codes over Eisenstein integers equipped with the hexagonal metric are more efficient, that is less power hungry and longer lasting. The Euclidean and hexagonal distances of two signal points $\alpha,\theta\in \mathcal{E}_{\eta}$ are  
\begin{align*}
    d_{\rm E}(\alpha,\theta)&=\|\alpha-\theta\|=\sqrt{N_{\rho}(\alpha-\theta)} \mbox{ and}\\
    d_{\rm Hex}(\alpha,\theta)&=\wt_{\rm Hex}(\alpha-\theta).
\end{align*}
The minimum Euclidean and hexagonal distances of the constellation are, therefore,
\begin{align*}
    d_{\rm E}(\mathcal{E}_{\eta}) &=\min_{\alpha,\theta\in \mathcal{E}_{\eta},\;\alpha\neq\theta} d_{\rm E}(\alpha,\theta) \mbox{ and}\\
    d_{\rm Hex}(\mathcal{E}_{\eta}) &=\min_{\alpha,\theta\in \mathcal{E}_{\eta},\;\alpha\neq\theta} d_{\rm Hex}(\alpha,\theta).
\end{align*}

\subsection{Set Partitioning on Codes over Eisenstein Integers} 

The minimum squared Euclidean distance in the Eisenstein constellations is $d^{2}_{\rm E}(\mathcal{E}_{\eta})=1$. The proposed construction allows for a natural division into additive subgroups and their cosets. Doing this may increase the minimum distances and simplify their exact determination. We begin our treatment with a corollary from a known result. 

\begin{theorem}\cite{Stern2019}\label{addsubgrup}
Let $A$ be an additive group with respect to the addition $\mu_{\eta}(\cdot )$. If $A'=c+A$ such that $c\in \Z[\rho]\setminus A$ is any coset of $A$, then 
\begin{equation*}
    d^{2}_{\rm E}(A)=d^{2}_{\rm E}(A')=\min_{\alpha\in A\setminus\{0\}}N_{\rho}(\alpha).
\end{equation*}
\end{theorem}

\begin{corollary}
Let $\mathcal{E}_{\eta}$ be an additive group with respect to the addition $\mu_{\eta}(\cdot)$. If $\mathcal{E}_{\eta}' = c+\mathcal{E}_{\eta}$ with $c\in \Z[\rho]\setminus \mathcal{E}_{\eta}$ is any coset of $\mathcal{E}_{\eta}$, then 
\begin{equation*}
    d_{\rm Hex}(\mathcal{E}_{\eta})=d_{\rm Hex}(\mathcal{E}_{\eta}')=\min_{\alpha\in \mathcal{E}_{\eta}\setminus\{0\}}\wt_{\rm Hex}(\alpha).
\end{equation*}
\end{corollary}
It is clear, therefore, that to obtain a larger minimum distance than that of the original set, we partition $R_{\eta}$ into equal-sized subsets by determining a subgroup $H$ of $\Z[\rho]/\langle\eta\rangle$ and its cosets
\[
H=H^{(0)}, H^{(1)},\ldots, H^{(N_{\rho}(\eta)/|H|-1)}.
\]
They correspond to $\mathcal{E}_{\eta}^{(0)}, \mathcal{E}_{\eta}^{(1)},\ldots,\mathcal{E}_{\eta}^{(N_{\rho}(\eta)/|H|-1)}$, each with $|H|$ elements. We avoid any subgroup that contains some units in $\Z[\rho]$ since it has minimum distance $1$.

\begin{example}
Let $\eta\sim n\in \N$, that is, $\gcd(a,b)=n$. Hence,  $\Z[\rho]/\langle\eta\rangle\cong \Z[\rho]/\langle n\rangle\cong \Z_{n}[\rho]$ of order $n^2$. We observe that 
 \begin{align*}
    H_{1}&:=\{a+0\rho\,:\, a\in\Z_{n}\}\subset\Z_{n}[\rho] \mbox{ and}\\
     H_{2}&:=\{0+b\rho\,:\, b\in\Z_{n}\}\subset\Z_{n}[\rho]
\end{align*}
are subgroups of $\Z_{n}[\rho]$ of order $n$. The respective cosets of $H_{1}$ and $H_{2}$ are 
\begin{align*}
H_{1}^{(k)} &=k\rho+H_{1}= \{k\rho+z\,:\, z\in H_{1}\} \\
&=\{a+k\rho\,:\, a\in \Z_{n}\} \mbox{ and}\\
H_{2}^{(k)} &=k+H_{2}= \{k+z\,:\, z\in H_{2}\}\\
&=\{k+b\rho\,:\, b\in \Z_{n}\},
\end{align*}
for each $k \in \{0,1,\ldots,n-1\}$.
The subsets correspond to the respective sets $\mathcal{E}_{n}^{(0)}, \mathcal{E}_{n}^{(1)},\ldots,\mathcal{E}_{n}^{(n-1)}$, each with $n$ elements, but the minimum distance remains $1$. 
\end{example}

\begin{example}
For $\eta=a+b\rho=t(c+d \, \rho)$ with $\gcd(a,b)=t$, $1<t<n$, and $\gcd(c,d)=1$, Theorem \ref{classequivringfac} tells us that
\begin{align*}
L_{1} &:=\{[x+ 0 \, \rho]\,:\, 0\leq x< t \, N_{\rho}(c+ d \, \rho)\} \mbox{ and}\\
L_{2}&:=\{[0+y\rho]\,:\, 0\leq y<t\}
\end{align*} 
are subgroups of 
\begin{multline*}
\Z[\rho]/\langle\eta\rangle=\{[x+y\rho]\,: 0\leq x<t \, N_{\rho}(m+n\, \rho) \\\mbox{ and } 0\leq y<t\}.
\end{multline*} 
The minimum distance is $1$. 
\end{example} 

Our next result describes how to find additive subgroups that lead to a minimum distance $ >1$.
\begin{theorem}
Let $\eta=a+b\rho=t(m+n\rho)\in \Z[\rho]\setminus\{0\}$ with $t=\gcd(a,b)$.
\begin{enumerate}
\item[i)] If $t=1$ and $N_{\rho}(\eta)=cd$ is a composite number, then 
\begin{equation}\label{subggcd1}
H_{1}=\{[x]_{\eta}\in\Z[\rho]/\langle\eta\rangle\,:\, x \in \{0,c,\ldots,(d-1)c\}\}  
\end{equation} 
is a subgroup of $\Z[\rho]/\langle\eta\rangle$ with minimum distance $c$.
\item[ii)] If $t = cd>1 $ for some $c,d\in \N$, then 
\begin{align}\label{subg}
& H_{2}=\{[x+ y \, \rho]_{\eta}\in\Z[\rho]/\langle\eta\rangle \,: \notag \\
&\quad x \in \{0,c,\ldots, (dN_{\rho}(m+n\rho)-1) c \} \notag \\
&\quad \mbox{ and } y \in \{0,c,\ldots,(d-1)c\}\}  
\end{align} 
is a subgroup of $\Z[\rho]/\langle\eta\rangle$ with minimum distance $c$.
\end{enumerate}
The cosets of $H_i$ correspond to $\mathcal{E}_{\eta}^{(0)}, \mathcal{E}_{\eta}^{(1)},\ldots,\mathcal{E}_{\eta}^{(N_{\rho}(\eta)/|H_{i}|-1)}$, each with $|H_{i}|$ elements whose minimum distance is $c$.
\end{theorem}
\begin{IEEEproof}
To verify the first assertion, let $\alpha_{1},\alpha_{2}\in H_{1}$. Then $\alpha_{1}=x_{1}$ and $\alpha_{2}=x_{2}$ for some $x_{1},x_{2}\in\{0,c,\ldots,
(d-1)c\}$. By Theorem \ref{classequivringfac} and since $x_{1}-x_{2}$ is multiple of $c$, we have $\alpha_{1}-\alpha_{2}=x_{1}-x_{2}\in H_{1}$, confirming that the distance of two distinct elements of $H_{1}$ is $\geq c$.

For the second assertion, letting $\alpha_{1},\alpha_{2}\in H_{2}$, we write $\alpha_{1}=x_{1}+y_{1}\rho$ and $\alpha_{2}=x_{2}+y_{2}\rho$ for some $x_{1},x_{2}\in\{0,c,\ldots,
(dN_{\rho}(m+n\rho)-1) c\}$ and $y_{1},y_{2}\in\{0,c,\ldots,
(d-1)c\}$. By Theorem \ref{classequivringfac} and since both $x_{1}-x_{2}$ and $y_{1}-y_{2}$ are multiple of $c$, we have 
\[
\alpha_{1}-\alpha_{2}=(x_{1}-x_{2})+(y_{1}-y_{2})\rho\in H_{2}.
\]
The distance of two distinct elements of $H_{2}$ is, therefore, $\geq c$. 

Since $\alpha\equiv \mu_{\eta}(\alpha)\Mod{\eta}$ for every $\alpha\in \mathcal{R}_{\eta}$, the cosets of $\mathcal{E}_{\eta}^{(0)}$ have minimum distance $c$. 
\end{IEEEproof}

The next two corollaries can be inferred immediately.

\begin{corollary}
Let $\eta$ be a primitive Eisenstein integer. The ring $\mathcal{E}_{\eta}$ of size $N_{\rho}(\eta)=c_{1}c_{2} \cdots c_{s}$ can be partitioned as follows.
\begin{enumerate}
    \item Partition $\mathcal{E}_{\eta}$ into $c_{1}$ subsets of minimum squared Euclidean distance $c_{1}$.
    \item Partition each of the $c_1$ subsets into $c_{2}$ subsets of minimum squared Euclidean distance $c_{1}c_{2}$.
    \item Continue the process until each of the $c_{s-2}$ subsets is partitioned into $c_{s-1}$ subsets of minimum squared Euclidean distance $c_{1}c_{2} \cdots c_{s-1}$.
\end{enumerate}
\end{corollary}

\begin{figure}[ht]
    \centering
    \includegraphics[width=0.9\linewidth]{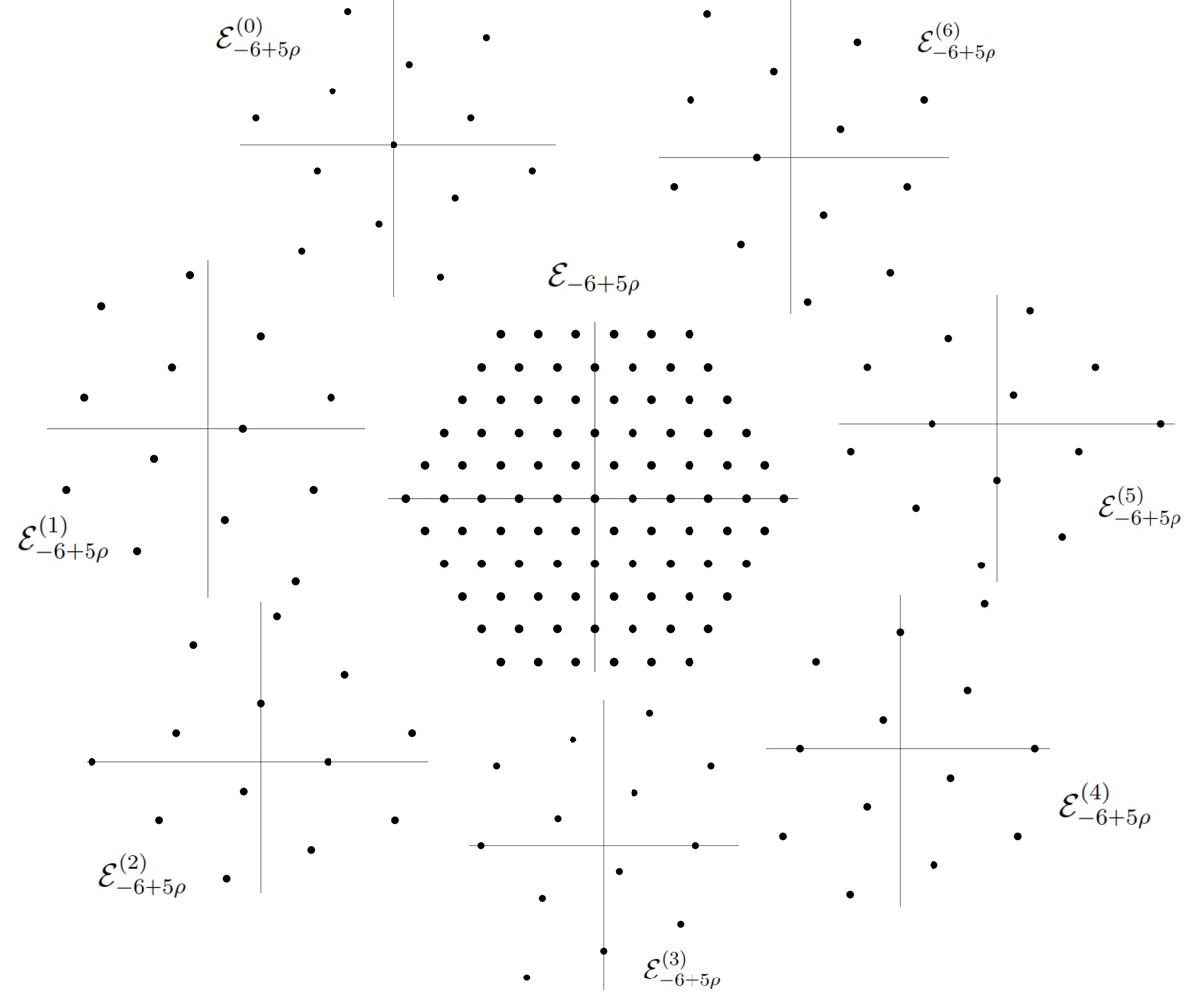}
    \caption{Eisenstein constellation $\mathcal{E}_{(-6+5\rho)}$ and its partitions.}
    \label{fig:SPpsi1psi2}
\end{figure}

\begin{example}
Given primes $\psi_{1}=2+3\rho$ and $\psi_{2}=3+4\rho$, we have $\eta=\psi_{1}\psi_{2}=-6+5\rho$ and $\mathcal{E}_{\psi_{1}\psi_{2}}$ being isomorphic to $\Z_{91}$, since $91=N_{\rho}(\psi_{1}\psi_{2})$. We can partition $\mathcal{E}_{\psi_{1}\psi_{2}}$ into $7$ or $13$ subsets with minimum squared Euclidean distance $7$ or $13$, accordingly. If into $7$ subsets, then they correspond to
\begin{align*}
  \Z_{91}^{(0)}&=\{0,7,14,21,28,35,42,49,56,63,70,77,84\},\\
\Z_{91}^{(1)}&=\{1,8,15,22,29,36,43,50,57,64,71,78,85\},\\
   \Z_{91}^{(2)}&=\{2,9,16,23,30,37,44,51,58,65,72,79,86\},\\
   \Z_{91}^{(3)}&=\{3,10,17,24,31,38,45,52,59,66,73,80,87\},\\
  \Z_{91}^{(4)}&=\{4,11,18,25,32,39,46,53,60,67,74,81,88\},\\
    \Z_{91}^{(5)}&=\{5,12,19,26,33,40,47,54,61,68,75,81,89\},\\
    \Z_{91}^{(6)}&=\{6,13,20,27,34,41,48,55,62,69,76,83,90\}.
    \end{align*}
Considering $z\in \Z_{91}^{(0)}\setminus\{0\}=\{7,14,21,\ldots,84\}$, the minimum value of the norm is not determined by $\mu_{\eta}(7)$, since $N_{\rho}(\mu_{\eta}(7))=21$, but by $\mu_{\eta}(14)$, since $N_{\rho}(\mu_{\eta}(14))=7$. The subsets have minimum squared Euclidean distance $7$. The set $\Z_{91}^{(0)}$ forms an additive subgroup of $\Z_{49}$, whereas $\Z_{91}^{(1)},\ldots,\Z_{91}^{(6)}$ are cosets of $\Z_{91}^{(0)}$. The minimum squared Euclidean distance in the subsets is determined by the minimum  Euclidean norm of the nonzero elements of $\Z_{91}^{(0)}$. Figure \ref{fig:SPpsi1psi2} provides the visualization.
\end{example}

\begin{corollary}
Let $\eta=a+b\rho=t(m+n\rho)$, with $t=\gcd(a,b)=c_{1}c_{2}\cdots c_{s}>1$, be a nonprimitive Eisenstein integer. The ring $\mathcal{E}_{\eta}$ of size $N_{\rho}(\eta)=t^{2}N_{\rho}(m+n\rho)$ can be recursively partitioned in the following manner. 
\begin{enumerate}
    \item Partition $\mathcal{E}_{\eta}$ into $c_{1}$ subsets of minimum squared Euclidean distance $c_{1}$.
    \item Partition each of the $c_1$ subsets into $c_{2}$ subsets of minimum squared Euclidean distance $c_{1}c_{2}$.
    \item Continue the process until each of the $c_{s-2}$ subsets is partitioned into $c_{s-1}$ subsets of minimum squared Euclidean distance $c_{1}c_{2} \cdots c_{s-1}$.
\end{enumerate}
\end{corollary}

\begin{example}
For Eisenstein primes $\psi_{1}=2+3\rho,\;\psi_{2}=3+4\rho$, and $\beta=1-\rho$, we have $\mathcal{E}_{\beta\psi_{1}\psi_{2}} \sim \Z_{273}$ since $273=N_{\rho}(\beta\psi_{1}\psi_{2})$. We can partition
\[
\mathcal{E}_{\beta\psi_{1}\psi_{2}}=\{z\Mod{-1+16\rho}\,:\, z\in \Z_{273}\}.
\]
into $7$ subsets, each with minimum squared Euclidean distance $7$, corresponding to the integer sets 
\begin{align*}
\Z_{273}^{(0)}&=\{0,7,14,\ldots,266\}, & \Z_{273}^{(4)}&=\{4,11,18,\ldots,270\},\\
\Z_{273}^{(1)}&=\{1,8,15,\ldots,267\}, &\Z_{273}^{(5)}&=\{5,12,19,\ldots,271\},\\
\Z_{273}^{(2)}&=\{2,9,16,\ldots,268\}, & \Z_{273}^{(6)}&=\{6,13,20,\ldots,272\}.\\
\Z_{273}^{(3)}&=\{3,10,17,\ldots,269\}.& &
\end{align*}
Each of these subsets can be partitioned into $13$ subsets, each having minimum squared Euclidean distance $91$. We partition $\Z_{273}^{(0)}$ into
\begin{align*}
\Z_{273}^{(0,0)} &=\{0,91,182\},& \Z_{273}^{(0,7)}&=\{49,140,231\},\\
\Z_{273}^{(0,1)} &=\{7,98,189\},&\Z_{273}^{(0,8)}&=\{56,147,238\},\\
\Z_{273}^{(0,2)} &=\{14,105,196\},& \Z_{273}^{(0,9)}&=\{63,154,245\},\\
\Z_{273}^{(0,3)} &=\{21,112,203\},&\Z_{273}^{(0,10)}&=\{70,161,252\},\\
\Z_{273}^{(0,4)} &=\{28,119,210\},& \Z_{273}^{(0,11)}&=\{77,168,259\},\\
\Z_{273}^{(0,5)} &=\{35,126,217\},&\Z_{273}^{(0,12)}&=\{84,175,266\}.\\
\Z_{273}^{(0,6)} &=\{42,133,224\}.&&   
\end{align*}
The sets $\displaystyle{\mathcal{E}_{\beta\psi_{1}\psi_{2}},\;\mathcal{E}_{\beta\psi_{1}\psi_{2}}^{(0)},\;\mathcal{E}_{\beta\psi_{1}\psi_{2}}^{(0,0)}}$ correspond, respectively, to $\Z_{273}$, $\Z_{273}^{(0)}$, and $\Z_{273}^{(0,0)}$.
\end{example}

\begin{example}
The set 
\[
S_{1}=\{x+y\rho\;:\;x,y\in\{0,2,4\}\}\subset \Z_{6}[\rho]
\]
is a subgroup of $\Z_{6}[\rho]$. Its cosets $S_{1}^{(0)}=S_{1}$, $S_{1}^{(1)}=1+S_{1}$, $S_{1}^{(2)}=\rho+S_{1}$, and $S_{1}^{(3)}=(1+\rho)+S_{1}$ correspond respectively to $\mathcal{E}_{6}^{(0)}$, $\mathcal{E}_{6}^{(1)}$, $\mathcal{E}_{6}^{(2)}$, and $\mathcal{E}_{6}^{(3)}$. The subsets, as shown in Figure \ref{fig:SPE6}, have minimum Euclidean distance $2$.
\begin{figure}[ht]
    \centering
    \includegraphics[width=0.9\linewidth]{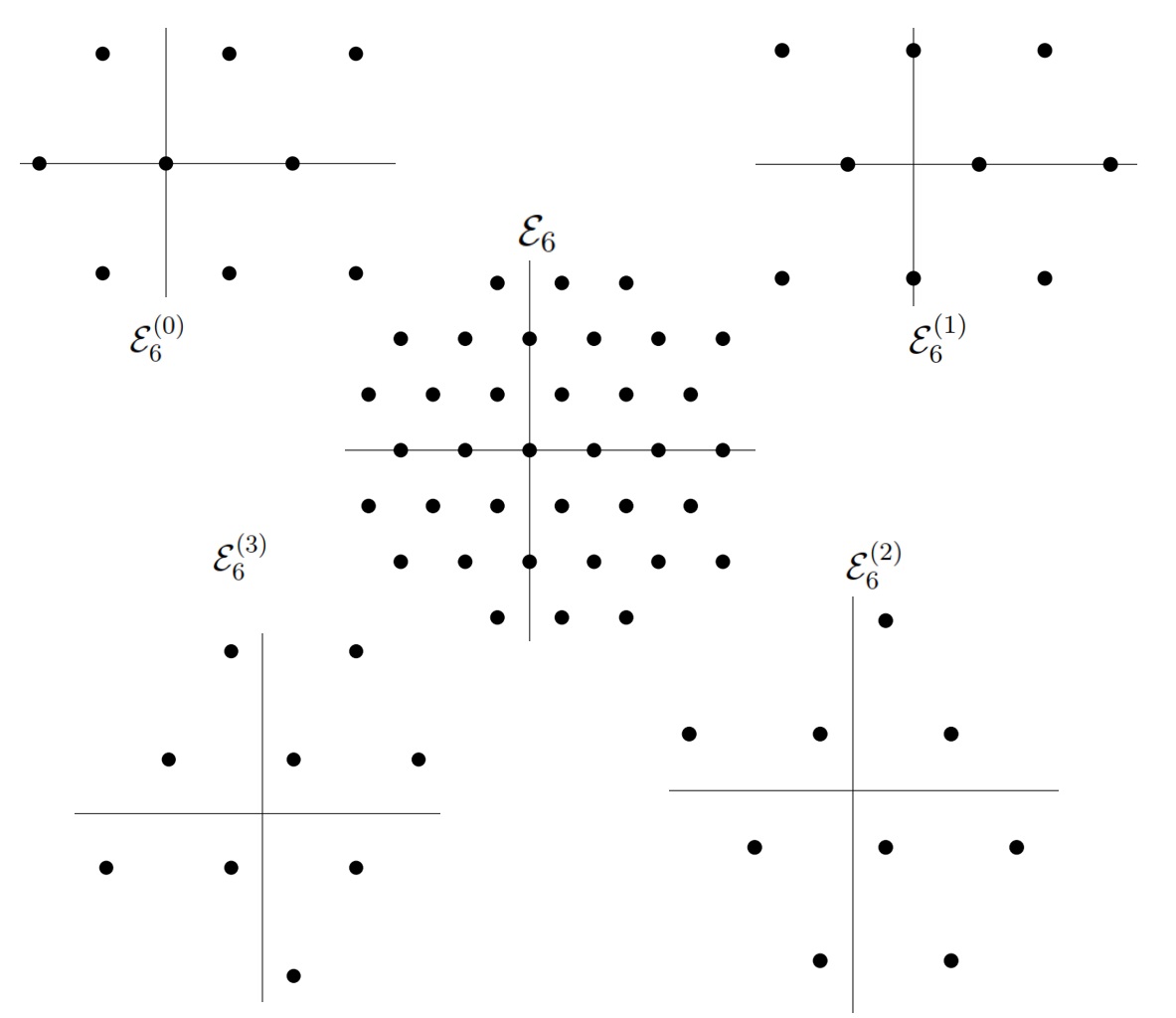}
    \caption{Eisenstein constellation $\mathcal{E}_{6}$ and its partitions.}
    \label{fig:SPE6}
\end{figure}
We can also partition $\mathcal{E}_{6}$ into $9$ subsets of minimum distance $3$ if so desired.
\end{example}

\begin{example}
For $\eta=6+12\rho=6(1+2\rho)$, we have 
\[
\Z[\rho]/\langle \eta\rangle=\{[x+y\rho]_{\eta}\;:\;0\leq x<18\;\;\mbox{and}\;\;0\leq y<6\}.
\]
The set 
\[
H_{1}=\{[x+ y \, \rho]_{\eta}~:~
x \in \{0,3,6,\ldots,15\} \mbox{ and } y \in \{0,3\}\}
\]
is a subgroup of $\Z[\rho]/\langle \eta\rangle$ whose respective cosets 
\begin{align*}
& H_{1}^{(0)}=H_{1}, ~ H_{1}^{(1)}=1+H_{1}, ~ H_{1}^{(2)} =2+H_{1},\\
& H_{1}^{(3)}=\rho+H_{1}, ~ H_{1}^{(4)} = 1+\rho+H_{1}, ~ H_{1}^{(5)}=2+\rho+H_{1},\\
& H_{1}^{(6)}=2\rho+H_{1}, ~ H_{1}^{(7)}=1+2\rho+H_{1}\mbox{, and}\\
&H_{1}^{(8)}=2+2\rho+H_{1}
\end{align*}
correspond to $\mathcal{E}_{6+12\rho}^{(0)}$, $\mathcal{E}_{6+12\rho}^{(1)}$, $\ldots$, $\mathcal{E}_{6+12\rho}^{(8)}$. Figure \ref{fig:SPE612rho} depicts the subsets, each with minimum Euclidean distance $3$.
\begin{figure}
    \centering
    \includegraphics[width=0.9\linewidth]{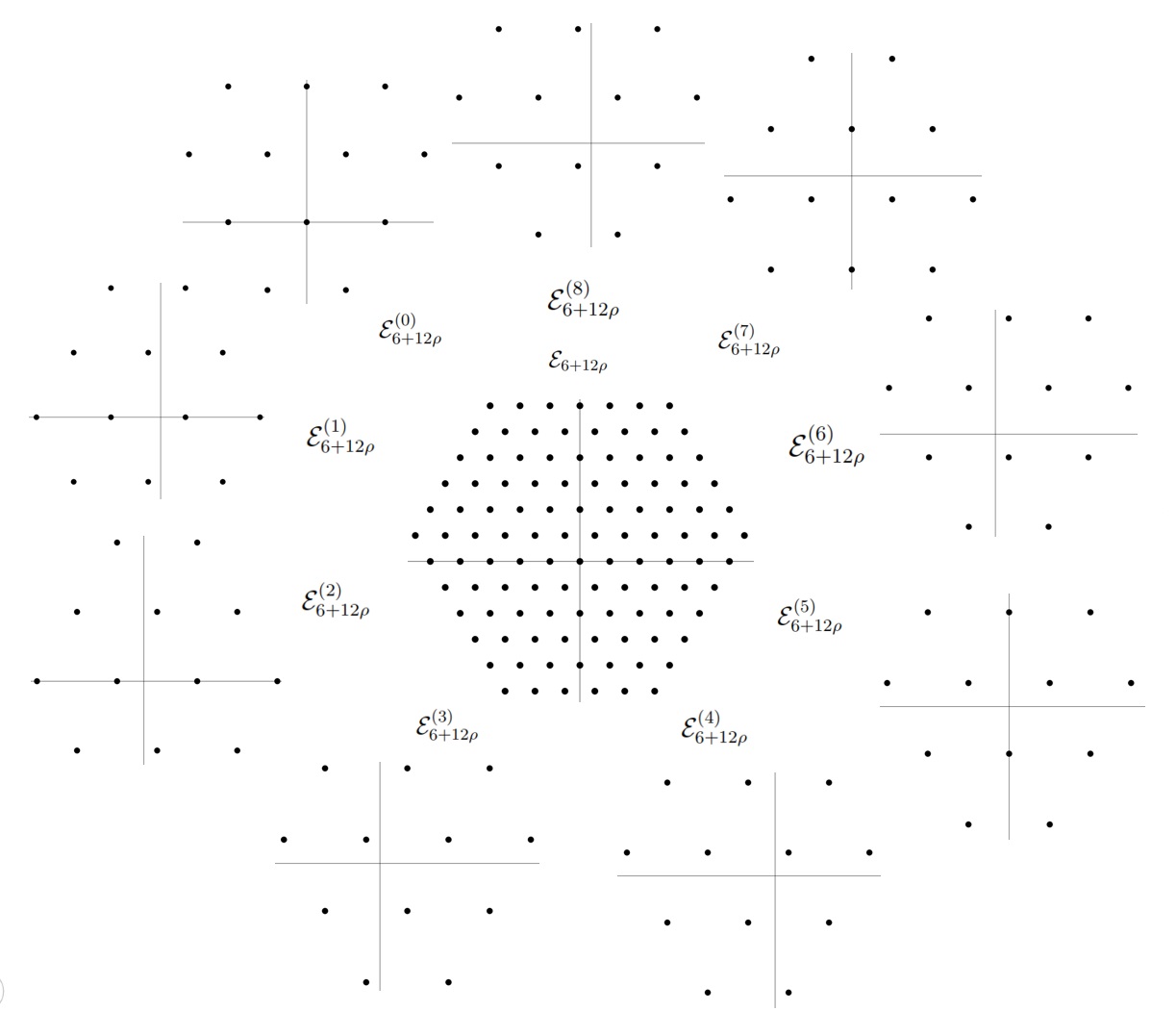}
    \caption{Eisenstein constellation $\mathcal{E}_{6+12\rho}$ and its partitions. }
    \label{fig:SPE612rho}
\end{figure}
\end{example}  

\subsection{Comparison between Codes over Eisenstein Integers and Gaussian Integers}

We compare codes over $\Z[i]$ and codes over $\Z[\rho]$ in terms of their average energies with respect to the Euclidean, square Euclidean, Mannheim, and hexagonal metrics. 

We recall from \cite{Dresden2005} that, for $\eta=a+bi=t(c+di)\in\Z[i]\setminus\{0\}$ with $t=\gcd(a,b)$ and $\gcd(c,d)=1$, the complete residue system is
\begin{align*}
    \Z[i]/\langle \eta\rangle=\{[x+yi]_{\eta} \, : \, 0\leq x<tN(c+di),\;0\leq y<t\},
\end{align*}
with $[x+yi]_{\eta}=x+yi+\langle\eta\rangle$. Letting 
\[
\mathcal{T}_{\eta}=\{x+yi \, : \, 0\leq x<tN(c+di),\;0\leq y<t\},
\] 
we get Gaussian constellation 
\[
\mathcal{G}_{\eta}=\{\alpha\Mod{\eta}\;:\;\alpha\in \mathcal{T}_{\eta}\},
\]
with 
\[\alpha\Mod{\eta}= \alpha-\lfloor\tfrac{\alpha\overbar{\eta}}{N(\eta)}\rceil \eta.
\]

We use the definition of the Mannheim distance in \cite{Martinez2007}. For brevity, we let the class $K(\theta)$ of a given Gaussian integer $\theta$ be the set of all Gaussian integers $\alpha$ such that $\theta=\alpha\Mod{\eta}$. The Mannheim weight of $\theta$ is
\[
\wt_{\rm M}(\theta)=\min_{a+bi\in K(\theta)}|a|+|b|.
\]

\begin{table*}[!ht]
\caption{Comparison among Gaussian and Eisenstein Constellations of the same cardinality in terms of the average energies with respect to the Euclidean, square Euclidean, Mannheim and hexagonal distances. These are denoted, respectively by $E$, $E^2$, $E_{\rm M}$, and $E_{\rm Hex}$.}
\label{compcodegausEissame}
\renewcommand{\arraystretch}{1.1}
\centering
\begin{tabular}{llc rrr rrr}
\toprule
Gaussian $\mathcal{G}_{\eta}$ & Eisenstein $\mathcal{E}_{\eta}$ &  Size &$E(\mathcal{G}_{\eta})$&$E(\mathcal{E}_{\eta})$& $E^{2}(\mathcal{G}_{\eta})$&$E^{2}(\mathcal{E}_{\eta})$&$E_{\rm M}(\mathcal{G}_{\eta})$&$E_{\rm Hex}(\mathcal{E}_{\eta})$\\
\midrule
$\mathcal{G}_{2}$  &$\mathcal{E}_{2}$& $4$ & $0.85$ & $0.75$ & $1.00$ & $0.75$ & $1.00$ & $0.75$ \\
$\mathcal{G}_{3}$  &$\mathcal{E}_{3}$& $9$ & $1.07$ & $1.05$ & $1.33$ & $1.33$ & $1.33$ & $1.11$\\
$\mathcal{G}_{2+3i}$  &$\mathcal{E}_{3+4\rho}$ & $13$ & $1.36$
 & $1.26$ & $2.15$ & $1.85$ & $1.54$ & $1.38$\\

$\mathcal{G}_{4}$  &$\mathcal{E}_{4}$ & $16$ & $1.59$ & $1.40$ & $3.00$ & $2.25$ & $2.00$ & $1.50$\\

$\mathcal{G}_{-3+4i}$  & $\mathcal{E}_{5+5\rho}$ & $25$ & $1.90$  & $1.77$ & $4.16$ & $3.60$ & $2.24$ & $1.92$ \\

$\mathcal{G}_{5}$  &$\mathcal{E}_{5}$ & $25$ & $1.87$ & $1.77$ & $4.00$ & $3.60$ & $2.40$ & $1.92$\\
$\mathcal{G}_{6}$  &$\mathcal{E}_{6}$& $36$ & $2.34$ & $2.11$ & $6.33$ & $5.08$ & $3.00$ & $2.31$\\
$\mathcal{G}_{6+i}$ &$\mathcal{E}_{7+3\rho}$  & $37$ & $2.32$ & $2.11$ & $6.16$ & $5.03$ & $2.92$ & $2.27$ \\
$\mathcal{G}_{7}$  &$\mathcal{E}_{7}$& $49$ & $2.70$ & $2.46$ & $8.29$ & $6.86$ & $3.51$ & $2.69$\\
$\mathcal{G}_{8}$  &$\mathcal{E}_{8}$& $64$ & $3.09$ & $2.82$ & $11.00$ & $9.00$ & $4.00$ & $3.09$\\
$\mathcal{G}_{8+3i}$  &$\mathcal{E}_{8+9\rho}$ & $73$ & $3.27$ & $2.99$ & $12.16$ & $10.11$ & $4.16$ & $3.29$\\
$\mathcal{G}_{9}$  &$\mathcal{E}_{9}$& $81$ & $3.55$ & $3.17$ & $14.31$ & $11.33$ & $4.65$ & $3.48$\\
$\mathcal{G}_{10}$ & $\mathcal{E}_{10}$& $100$ & $3.85$ & $3.51$ & $17.00$ & $13.95$ & $5.00$ & $3.87$\\
$\mathcal{G}_{11}$  &$\mathcal{E}_{11}$& $121$ & $4.19$ & $3.87$ & $20.00$ & $16.91$ & $5.45$ & $4.26$\\
$\mathcal{G}_{12}$  &$\mathcal{E}_{12}$& $144$ & $4.61$ & $4.22$ & $24.33$ & $20.08$ & $6.00$ & $4.65$ \\
$\mathcal{G}_{13}$  &$\mathcal{E}_{13}$& $169$ & $4.96$ & $4.57$ & $28.00$ & $23.54$ & $6.46$ & $5.04$\\
$\mathcal{G}_{-5+12i}$  &$\mathcal{E}_{-7+8\rho}$ & $169$ & $4.97$ & $4.55$ & $28.17$ & $23.36$ & $6.30$ & $4.97$\\
$\mathcal{G}_{14}$  &$\mathcal{E}_{14}$& $196$ & $5.38$ & $4.92$ & $33.00$ & $27.32$ & $7.00$ & $5.43$\\
$\mathcal{G}_{8+12i}$  &$\mathcal{E}_{12+16\rho}$ & $208$ & $5.52$ & $5.06$ & $34.69$ & $28.90$ & $6.87$ & $5.57$\\
$\mathcal{G}_{15}$ &$\mathcal{E}_{15}$ & $225$ & $5.73$ & $5.27$ & $37.33$ & $31.33$ & $7.47$ & $5.82$ \\
$\mathcal{G}_{16}$  & $\mathcal{E}_{16}$ & $256$ & $6.14$ & $5.62$ & $43.00$ & $35.63$ & $8.00$ & $6.21$\\
$\mathcal{G}_{16+6i}$  &$\mathcal{E}_{16+18\rho}$& $292$ & $6.54$ & $6.00$ & $48.67$ & $40.57$ & $8.35$ & $6.63$ \\
$\mathcal{G}_{18+3i}$  &$\mathcal{E}_{21+9\rho}$& $333$ & $6.98$ & $6.40$ & $55.50$ & $46.25$ & $9.06$ & $7.00$\\
\bottomrule
\end{tabular}
\end{table*}

Table \ref{compcodegausEissame} shows that, when the alphabets are of the same cardinality, 
\[
E(\mathcal{E}_{\eta}) \leq E(\mathcal{G}_{\eta}), ~ E^{2}(\mathcal{E}_{\eta}) \leq E^{2}(\mathcal{G}_{\eta}), ~ E_{\rm Hex}(\mathcal{E}_{\eta})\leq E_{M}(\mathcal{G}_{\eta}).
\]
By Theorem \ref{relnormhex}, 
\[
E(\mathcal{E}_{\eta})\leq E_{\rm Hex}(\mathcal{E}_{\eta})\leq E^{2}(\mathcal{E}_{\eta}).
\]

\begin{figure}
    \centering
    \includegraphics[width=0.9\linewidth]{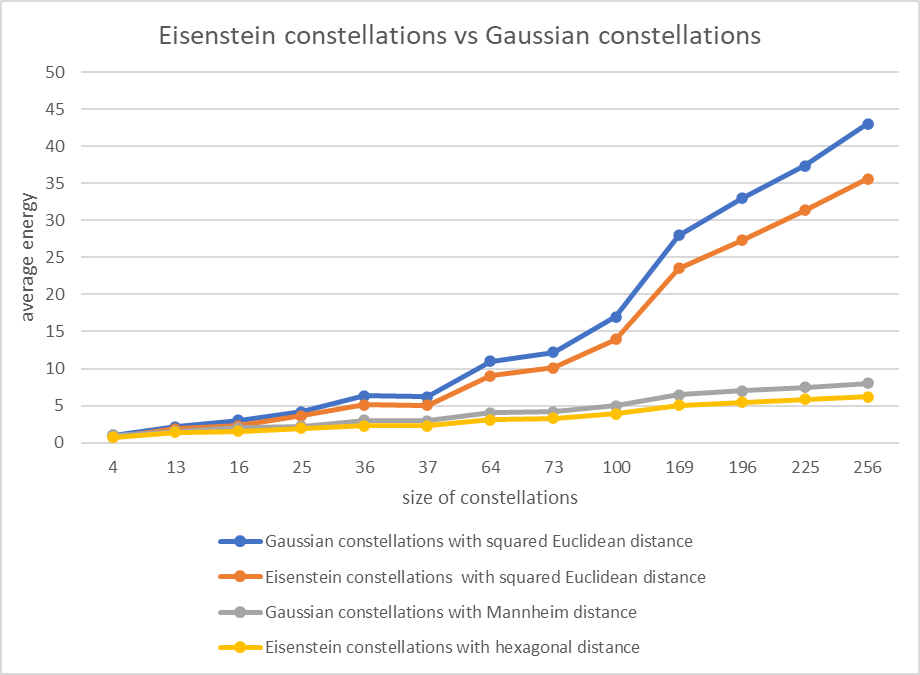}
    \caption{Comparison of the average energy among Eisenstein and Gaussian constellations.}
    \label{fig:ECvsGC}
\end{figure}

Figure \ref{fig:ECvsGC} visualizes the comparison of the average energies.

\section{Conclusion}\label{sec:conclu}

This work has introduced the concept of partitioning of Eisenstein integers based on additive subgroups. Our approach forms a powerful technique that leverages on the unique properties of Eisenstein integers to design efficient and robust coding schemes. These schemes can be applied in various communication systems to improve error correction and signal representation. It is natural to consider if our proposed set partitioning method for Eisenstein integers extends generally to the five imaginary quadratic Euclidean domains $\Z[\xi]$ for 
\[
\xi \in \left\{i,\sqrt{2}i,\rho,\frac{-1+\sqrt{7}i}{2},\frac{-1+\sqrt{11}i}{2}\right\}.
\]

\appendices
\section{Proof of Theorem \ref{classequivringfac}}
\begin{lemma}\cite{Ozkan2013}\label{lemsyrt}
Given integers $a$, $b$, $c$, $d$, and $k$, we have $c+d\rho\in \langle k(a+b\rho)\rangle$ if and only if $k(a^{2}+b^{2}-ab)$ divides both $((a-b)c+bd)$ and $(ad-bc)$.
\end{lemma}

Starting with the fact that 
$[x+y\rho]_{\eta}=[x'+y'\rho]_{\eta} $ if and only if $(x'-x)+(y'-y)\rho \in \langle \eta\rangle$, by Lemma \ref{lemsyrt} and computing in modulo $t \, N_{\rho}(m+n\rho)$, we have
\begin{align}
    (m-n)(x'-x)+n(y'-y) & \equiv 0 \mbox{ and } \label{eqr1a}\\
m(y'-y)-n(x'-x) &\equiv 0.\label{eqr2a}
\end{align}
Eliminating $x'-x$, we arrive at 
\begin{equation*}
(y'-y) \, N_{\rho}(m+n\rho)\equiv 0\Mod{tN_{\rho}(m+n\rho)},
\end{equation*} 
forcing $y'\equiv y\Mod{t}$.

We know from (\ref{eqr2a}) that 
\begin{equation}\label{eqr3}
n(x'-x)\equiv  m(y'-y)\Mod{tN_{\rho}(m+n\rho)}. 
\end{equation}
Subtracting (\ref{eqr2a}) from (\ref{eqr1a}) yields 
\begin{equation}\label{eqr4}
    m(x'-x)\equiv (n-m)(y-y')\Mod{tN_{\rho}(m+n\rho)}.
\end{equation} 
Assuming that $0\leq x'<tN_{\rho}(m+n\rho)$ and $0\leq y'<t$, if $\gcd(m,t)=1$ or $\gcd(n,t)=1$, then
\begin{equation*}
 \gcd(m,tN_{\rho}(m+n\rho))=1\,\text{or}\,\gcd(n,tN_{\rho}(m+n\rho))=1   
\end{equation*} 
due to $\gcd(m,N_{\rho}(m+n\rho))=1=\gcd(n,N_{\rho}(m+n\rho))$. If $0\leq y'<t$ and $y'\equiv y\Mod{t}$, then $y'=y\Mod{t}$. 
We observe further that, if $\gcd(n,tN_{\rho}(m+n\rho))=1$, then, by (\ref{eqr3}), $x' \equiv x-mn^{-1}(y-y')\Mod{tN_{\rho}(m+n\rho)}$. Since $0\leq x'<tN_{\rho}(m+n\rho)$, we know that 
\begin{equation*}
x'=x-mn^{-1}(y-y')\Mod{tN_{\rho}(m+n\rho)}.
\end{equation*}
\par If $\gcd(m,tN_{\rho}(m+n\rho))=1$, then, by (\ref{eqr4}),
\begin{equation*}
x'\equiv x+(m^{-1}n-1)(y-y')\Mod{tN_{\rho}(m+n\rho)}.
\end{equation*}
Since $0\leq x'<tN_{\rho}(m+n\rho)$, we have
\begin{equation*}
x'=x+(m^{-1}n-1)(y-y')\Mod{tN_{\rho}(m+n\rho)}.
\end{equation*}
 
\section{Proof of Theorem \ref{psikua}}
We will use Lemmas \ref{normassoconj} to \ref{prime2rel} in our proof of Theorem \ref{psikua}.

\begin{lemma}\cite{Loefgren2022}\label{normassoconj}
If $\gamma_{1}$ and $\gamma_{2}$ are Eisenstein primes such that $N_{\rho}(\gamma_{1})=N_{\rho}(\gamma_{2})$, then $\gamma_{1}\sim \gamma_{2}$ or $\gamma_{1}\sim \overbar{\gamma_{2}}$. If $N_{\rho}(\gamma_{1})=3$, then $\gamma_{1}\sim 1-\rho$. If $N_{\rho}(\gamma_{1})=p^2$, with $p\equiv 2\Mod 3$, then $\gamma_{1}\sim \overbar{\gamma_{1}}$. If $N_{\rho}(\gamma_{1})=q\equiv 1\Mod 3$ is a prime, then $\gamma_{1}\not\sim \overbar{\gamma_{1}}$.
\end{lemma}

We generalize Lemma \ref{normassoconj} to the next lemma.
\begin{lemma}\label{normksam}
If $\gamma_{1}, \gamma_{2}$ are Eisenstein primes such that $N_{\rho}(\gamma_{1}^k)=N_{\rho}(\gamma_{2}^k)$, then $\gamma_{1}^k\sim \gamma_{2}^k$ or $\gamma_{1}^k\sim \overbar{\gamma_{2}}^k$.  
\end{lemma}

\begin{IEEEproof}
The assumption $N_{\rho}(\gamma_{1}^k)=N_{\rho}(\gamma_{2}^k)$ implies 
\[
(N_{\rho}(\gamma_{1}))^{k}=(\gamma_{1}\overbar{\gamma_{1}})^k=(\gamma_{2}\overbar{\gamma_{2}})^k=(N_{\rho}(\gamma_{2}))^{k},
\]
which leads to 
\[
N_{\rho}(\gamma_{1})=\gamma_{1}\overbar{\gamma_{1}} = \gamma_{2}\overbar{\gamma_{2}}=N_{\rho}(\gamma_{2}).
\]
Since $\gamma_{1}$ is a prime, it must divide either $\gamma_{2}$ or $\overbar{\gamma_{2}}$. We prove the first case since the second one can be established analogously. If $\gamma_1 \mid \gamma_2$, then $\gamma_{2}=\theta \, \gamma_{1}$ for some $\theta\in \Z[\rho]$. We know, however, that $\gamma_{2}$ is a prime, forcing $\theta$ to be a unit in $\Z[\rho]$. Hence, $\gamma_{1}\sim \gamma_{2}$ and, thus, $\gamma_{1}^k \sim \gamma_{2}^k$.
\end{IEEEproof}

We know that if $\alpha \sim \theta $ then $N_{\rho}(\alpha )=N_{\rho}(\theta )$ over $\mathbb{Z}$. The converse, however, does not hold in general. Both $2+5\rho$ and $ -3-5\rho$ has norm $19$, for example, but they are not associates. We now provide a necessary and sufficient condition for two Eisenstein integers to have the same norm.

\begin{lemma}\label{normassosiate}
Elements $\alpha,\theta \in \mathbb{Z}[\rho ]$ have $N_{\rho}(\alpha )=N_{\rho}(\theta ) \in \mathbb{Z}$ if and only if $\alpha \sim \theta $ or $\alpha \sim \overbar{\theta}$. 
\end{lemma}
\begin{IEEEproof}
If $\alpha \sim \theta $, then $N_{\rho}(\alpha )=N_{\rho}(\theta )$. If $\alpha \sim \overbar{\theta}$, then $\alpha = \delta \, \overbar{\theta}$ for some unit $\delta\in \Z[\rho]$. Hence, 
\[
N_{\rho}(\alpha)= N_{\rho}(\delta \, \overbar{\theta}) = N_{\rho}(\delta) \, N_{\rho}(\overbar{\theta})=1 \, N_{\rho}(\overbar{\theta})=N_{\rho}(\theta).
\]

Conversely, if $N_{\rho}(\alpha )=N_{\rho}(\theta )$, then we factor $\alpha$ and $\theta$ as
\[
\alpha=u_{1} \, \gamma_{1}^{a_{1}} \, \gamma_{2}{}^{a_{2}} \ldots \gamma_{k}^{a_{k}} \mbox{ and }
\theta = u_{2} \, \gamma_{1}^{b_{1}} \, \gamma_{2}^{b_{2}} \ldots \gamma_{k}^{b_{k}},
\]
where $u_{1}$ and $u_{2}$ are units and, for each $i \in \{1,2,\ldots,k\}$, $\gamma_{i}$ is a prime in $\Z[\rho]$ and $a_i$ and $b_i$ are nonnegative integers. 
Since $N_{\rho}(\alpha)=N_{\rho}(\theta)$, we have $N_{\rho}(\gamma_{i}^{a_i}) = N_{\rho}(\gamma_{i}^{b_i})$, implying $a_i=b_i$ for all $i$. By Theorem \ref{normksam}, $\gamma_i{}^{a_i}\sim \gamma_i{}^{a_i}$ or $\gamma_i{}^{a_i}\sim \overbar{\gamma_i}{}^{a_i}$, forcing $\alpha \sim \theta $ or $\alpha \sim \overbar{\theta}$. 
\end{IEEEproof}

\begin{lemma}\cite{Gullerud2020}\label{prime2rel}
Let $ \psi=x+y\rho $ be a prime in $ \Z[\rho]$, with $ N_{\rho}(\psi)=\psi\overbar{\psi}=q\equiv 1\Mod 3 $ being a prime in $\Z$. If $ \psi^{n}=a+b\rho, $ then $ \gcd(a,b)=1 $ for all $ n\in \N$.
\end{lemma}

We proceed to proving Theorem \ref{psikua} by induction on $m$. Lemma \ref{prime2rel} ensures that the claim holds for $m=1$. Let $\psi_{1}^{r_{1}} \, \cdots \, \psi_{k}^{r_{k}}=c+ d \, \rho$, where $\gcd(c,d)=1$ for some $k\geq
 1$. Considering $\psi_{1}^{r_{1}}\cdots\psi_{k+1}^{r_{k+1}}$, we observe that 
 \begin{align*}
\psi_{1}^{r_{1}} \cdots \psi_{k+1}^{r_{k+1}} & = \psi_{1}^{r_{1}} \cdots \psi_{k}^{r_{k}} \, \psi_{k+1}^{r_{k+1}} =(c+d \, \rho)(x+y \, \rho)\\
&=(xc-yd)+(xd+yc-yd) \, \rho.
 \end{align*}
Let $P=xc-yd$ and $Q=xd+yc-yd$. It suffices to show that $\gcd(P,Q)=1$. If $\gcd(P,Q)=r\geq1$, then $r$ divides both $P$ and $Q$. Since $q_{1}^{r_{1}}\cdots q_{k+1}^{r_{k+1}}=N_{\rho}(\psi_{1}^{r_{1}}\cdots\psi_{k+1}^{r_{k+1}})=P^2+Q^2-PQ$, $r$ divides $P^2+Q^2-PQ$ and, hence, $r$ divides $q_{1}^{r_{1}}\cdots q_{k+1}^{r_{k+1}}$. Since the $q_{i}$s are prime integers, $r=1$ or $r=q_{1}^{t_{1}}\cdots q_{k+1}^{t_{k+1}}$, with $0\leq t_{i}\leq r_{i}$. We now show that $r > 1$ is impossible. 

If $r=q_{1}^{t_{1}}\cdots q_{k+1}^{t_{k+1}}$ and $r > 1$, then $P=q_{1}^{t_{1}}\cdots q_{k+1}^{t_{k+1}}s_{1}$ and $Q=q_{1}^{t_{1}}\cdots q_{k+1}^{t_{k+1}}s_{2}$ for some $s_{1},s_{2}\in \Z$. Observe that 
\begin{multline*}
q_{1}^{r_{1}} \cdots q_{k+1}^{r_{k+1}}=
P^2+Q^2-PQ = \\ q_{1}^{2t_{1}}\cdots q_{k+1}^{2t_{k+1}}N_{\rho}(s_{1}+s_{2}\rho).
\end{multline*}
If $r_{i}=2t_{i}$ for all $i\in\{1,\ldots,k+1\}$, then $N_{\rho}(s_{1}+s_{2}\rho)=1$, implying that $s_{1}+s_{2}\rho$ is a unit. Hence, 
\begin{align*}
q_{1}^{t_{1}}\cdots q_{k+1}^{t_{k+1}}(s_{1}+s_{2}\rho)&\sim q_{1}^{t_{1}}\cdots q_{k+1}^{t_{k+1}},\\
P+Q\rho&\sim \psi_{1}^{t_{1}}\overbar{\psi_{1}^{t_{1}}}\cdots \psi_{k+1}^{t_{k+1}}\overbar{\psi_{k+1}^{t_{k+1}}},\\
\psi_{1}^{r_{1}}\cdots \psi_{k+1}^{r_{k+1}}&\sim \psi_{1}^{t_{1}}\overbar{\psi_{1}^{t_{1}}}\cdots \psi_{k+1}^{t_{k+1}}\overbar{\psi_{k+1}^{t_{k+1}}},\\
\psi_{1}^{t_{1}}\cdots \psi_{k+1}^{t_{k+1}}&\sim \overbar{\psi_{1}^{t_{1}}}\cdots\overbar{\psi_{k+1}^{t_{k+1}}}.
\end{align*}
This contradicts $\psi_{1}^{t_{1}}\cdots \psi_{k+1}^{t_{k+1}}\not\sim \overbar{\psi_{1}^{t_{1}}}\cdots\overbar{\psi_{k+1}^{t_{k+1}}}$, since $\psi_{i}\not\sim \overbar{\psi_{i}}$. 

Let $I$ be the largest nonempty subset of $J=\{1,\ldots,k+1\}$ such that $r_{i}\neq 2t_{i}$ for any $i \in I$. Partitioning $I$ into $I_{1}=\{i\in I \, : \,r_{i}>2t_{i}\}$ and $I_{2}=\{i \in I\,:\,r_{i}<2t_{i}\}$, we have 
\[
\prod_{i\in I_{1}}q_{i}^{r_{i}-2t_{i}} = \prod_{j\in I_{2}}q_{j}^{2t_{j}-r_{j}}N_{\rho}(s+t\rho).
\]
If $I_{2}$ is nonempty, then 
\[
\frac{\prod_{i\in I_{1}}q_{i}^{r_{i}-2t_{i}}} {\prod_{j\in I_{2}}q_{j}^{2t_{j}-r_{j}}} = N_{\rho}(s+t\rho),
\]
which contradicts $N_{\rho}(s_{1}+s_{2}\rho)\in\Z$ since the $q_{j}$s are distinct primes. 

If $I_{2}$ is the empty set, then
\[
\prod_{i\in I_{1}}q_{i}^{r_{i}-2t_{i}}= N_{\rho}(s_{1}+s_{2}\rho).
\]
By Lemma \ref{normassosiate}, 
\[
s_{1}+s_{2}\rho \sim \, \prod_{i\in I_{1}}\psi_{i}^{r_{i}-2t_{i}} \mbox{ or } \overbar{ \, \prod_{i\in I_{1}}\psi_{i}^{r_{i}-2t_{i}}}.
\]
If $s_{1}+s_{2}\rho\sim \prod_{i\in I_{1}}\psi_{i}^{r_{i}-2t_{i}}$, then 
\begin{align*}
q_{1}^{t_{1}}\cdots q_{k+1}^{t_{k+1}}(s_{1}+s_{2}\rho) &\sim q_{1}^{t_{1}}\cdots q_{k+1}^{t_{k+1}}\prod_{i\in I_{1}}\psi_{i}^{r_{i}-2t_{i}},\\
 P+Q\rho &\sim   \prod_{i\in J}\psi_{i}^{t_{i}}\overbar{\psi_{i}^{t_{i}}} \prod_{i\in I_{1}}\psi_{i}^{r_{i}-2t_{i}},\\
\prod_{i\in J}\psi_{i}^{r_{i}} & \sim  \prod_{i\in I_{1}}\psi_{i}^{r_{i}-t_{i}}\prod_{i\in J}\overbar{\psi_{i}^{t_{i}}}\prod_{i\in J\setminus I_{1}}\psi_{i}^{t_{i}},\\
 \prod_{i\in I_{1}}\psi_{i}^{t_{i}}\prod_{i\in J\setminus I_{1}}\psi_{i}^{r_{i}}&\sim \prod_{i\in J\setminus I_{1}}\psi_{i}^{t_{i}}\prod_{i\in J}\overbar{\psi_{i}^{t_{i}}},\\
 \prod_{i\in I_{1}}\psi_{i}^{t_{i}}\prod_{i\in J\setminus I_{1}}\psi_{i}^{r_{i}-t_{i}}& \sim \prod_{i\in J}\overbar{\psi_{i}^{t_{i}}},\\
  \prod_{i\in I_{1}}\psi_{i}^{t_{i}}\prod_{i\in J\setminus I_{1}}\psi_{i}^{t_{i}}& \sim \prod_{i\in J}\overbar{\psi_{i}^{t_{i}}},\\
    \prod_{i\in J}\psi_{i}^{t_{i}}& \sim \prod_{i\in J}\overbar{\psi_{i}^{t_{i}}},
\end{align*}
which is impossible since $\psi_{i}\not\sim \overbar{\psi_{i}}$.

If  $s_{1}+s_{2}\rho\sim \overbar{\prod_{i\in I_{1}}\psi_{i}^{r_{i}-2t_{i}}}$, then 
\[
\prod_{i\in I_{1}}\psi_{i}^{r_{i}-t_{i}}\prod_{i\in J\setminus I_{1}}\psi_{i}^{t_{i}} \sim 
\prod_{i\in I_{1}}\overbar{\psi_{i}^{r_{i}-t_{i}}}\prod_{i\in J\setminus I_{1}}\overbar{\psi_{i}^{t_{i}}},\]
which is also impossible. Thus, $r = \gcd(P,Q)=1$.

\section{Proof of Theorem \ref{betapsikua}}
\begin{theorem}\label{primebetarel}
Let $\beta=1-\rho$ and let $\psi=x+y\rho $ be a prime in $ \Z[\rho] $, with $ N_{\rho}(\psi)=\psi \, \overbar{\psi} = q\equiv 1\Mod 3 $ being a prime in $ \Z$. If $ \beta \, \psi^{n}=a+b\rho, $ then $ \gcd(a,b)=1 $ for all $ n\in \N$.
\end{theorem}

\begin{IEEEproof}
We prove by induction in $n$. If $n=1$, then 
\begin{align*}
\beta\psi^n &=\beta\psi=(1-\rho)(x+y\rho)=u+v\rho \mbox{ and }\\
N_{\rho}(\beta\psi) & =3q = u^2+v^2-uv.
\end{align*}
If we let $\gcd(u,v)=r \geq 1$, then $r \mid (u^2+v^2-uv = 3q)$. Since $3$ and $q$ are primes in $\Z$, we have $r \in \{1,3,q,3q\}$. 

To rule out $r > 1$ we write $u=rs$ and $v=rt$ for some $s,t\in \Z$ to get
\begin{equation}\label{eqbeta3q}
    3q=u^2+v^2-uv=r^2N_{\rho}(s+t\rho).
\end{equation}
We use (\ref{eqbeta3q}) to reach a contradiction in each case. If $r=3$, then $q=3N_{\rho}(s+t\rho)\equiv 0\Mod 3$, which contradicts $q \equiv 1\Mod 3$. If $r=q$, then $3=qN_{\rho}(s+t\rho)$, contradicting $qN_{\rho}(s+t\rho)>3$ due to $N_{\rho}(s+t\rho)\geq 1$ and $q\equiv 1\Mod 3$ being a prime in $\Z$. If $r=3q$, then $1=3qN_{\rho}(s+t\rho)$, which is absurd since both $3$ and $q$ are prime. Thus, $\gcd(u,v)=r=1$.

Assuming $\beta\psi^k=c+d\rho$, with $\gcd(c,d)=1$ for some $k \geq 1$, we get
\begin{multline*}
\beta \psi^{k+1} =\beta(x+y\rho)^{k+1} =\beta(x+y\rho)^{k}(x+y\rho)\\
=(c+d\rho)(x+y\rho) =(xc-yd)+(xd+yc-yd)\rho.
\end{multline*}

Let $A=xc-yd$ and $B=xd+yc-yd$. It suffices to prove that $r=\gcd(A,B)=1$. Since 
\[
3q^{k+1}=N_{\rho}(\beta\psi^{k+1})=A^2+B^2-AB,
\]
$r$ divides both $A^2+B^2-AB$ and $3q^{k+1}$. Since $3$ and $q$ are prime integers, $r\in\{1,3,3q,3q^2,\ldots,3q^{k+1},q,q^2,\ldots,q^{k+1}\}$. We are now ruling out $r > 1$. 

Writing $A=rs$ and $B=rt$ for some $s,t\in \Z$, we get
\begin{equation}\label{eq3q}
3q^{k+1}=A^2+B^2-AB=r^{2}N_{\rho}(s+t\rho).
\end{equation}
Let $r=q^m$ for $1 \leq m \leq k+1$. By (\ref{eq3q}), we arrive at $3q^{k+1}=q^{2m}N_{\rho}(s+t\rho)$ and consider three cases.
\begin{enumerate}
    \item If $2m>k+1$, then $q^{2m-(k+1)} \geq q>3$, contradicting $q^{2m-(k+1)}N_{\rho}(s+t\rho)=3$.
    \item If $2m<k+1$, then $N_{\rho}(s+t\rho)=3q^{k+1-2m} = N_{\rho}(\beta\psi^{k+1-2m})$. By Lemma \ref{normassosiate}, $s+t\rho\sim \beta\psi^{k+1-2m}$ or $\sim \overbar{\beta}\cdot\overbar{\psi}^{k+1-2m}$. Either way, we get $\psi \sim \overbar{\psi}$, contradicting $\psi\nsim\overbar{\psi}$.
    \item If $2m=k+1$, then $3q^{k+1}=q^{2m}N_{\rho}(s+t\rho)$, which implies $N_{\rho}(s+t\rho)=3$, associating $s+t\rho$ to $\beta$ or $\overbar{\beta}$. However, since $\beta\sim \overbar{\beta}$, we get 
    \[
    \beta\psi^{k+1}=A+B\rho=q^m(s+t\rho)\sim q^m\beta\sim q^m\overbar{\beta}.
    \]
    From $\psi^{k+1}\sim q^m=\psi^m\overbar{\psi}^m$ we derive $\psi\sim \overbar{\psi}$, contradicting $\psi\nsim\overbar{\psi}$.
\end{enumerate}

Let $r=3q^m$ for $0\leq m\leq k+1$. By (\ref{eq3q}), we get $q^{k+1}=3q^{2m}N_{\rho}(s+t\rho)$, which implies $q^{k+1}\equiv 0\Mod 3$. This is a contradiction since $q \equiv 1 \Mod 3$ being a prime forces $q^{k+1}\equiv 1\Mod 3$. Thus, $r=\gcd(A,B)=1$. 
\end{IEEEproof}

\medskip

We now proceed to prove Theorem \ref{betapsikua} by induction in $m$. Theorem \ref{primebetarel} settles the case of $m=1$.

Let $\beta\psi_{1}^{r_{1}}\cdots\psi_{k}^{r_{k}}=c+d\rho$, where $\gcd(c,d)=1$ for some $k\geq
 1$. We consider $\beta \, \psi_{1}^{r_{1}}\cdots\psi_{k+1}^{r_{k+1}}$ and observe that 
 \begin{align}
\beta \, \psi_{1}^{r_{1}}\cdots\psi_{k+1}^{r_{k+1}}&=\beta \, \psi_{1}^{r_{1}}\cdots\psi_{k}^{r_{k}} \, \psi_{k+1}^{r_{k+1}} \notag
     \\
     &=(c+d\rho)(x+y\rho) \notag \\
     &=(xc-yd)+(xd+yc-yd)\rho.
\end{align}
Let $P=xc-yd$ and $Q=xd+yc-yd$. It suffices to show that $r:=\gcd(P,Q)=1$. Since 
\[
3 \, q_{1}^{r_{1}}\cdots q_{k+1}^{r_{k+1}} = N_{\rho}(u \, \beta \, \psi_{1}^{r_{1}}\cdots \psi_{k+1}^{r_{k+1}}) = P^2+Q^2-PQ,
\]
we know that $r \mid (3 \, q_{1}^{r_{1}}\cdots q_{k+1}^{r_{k+1}})$. Since $3$ and $q_{i}$s are prime integers, $r=1$ or $r=q_{1}^{t_{1}}\cdots q_{k+1}^{t_{k+1}}$ or $r=3 \, q_{1}^{t_{1}}\cdots q_{k+1}^{t_{k+1}}$, with $0\leq t_{i}\leq r_{i}$. Next, we rule out $r > 1$. 

If $r=q_{1}^{t_{1}}\cdots q_{k+1}^{t_{k+1}} \neq 1$, then we can write
\[
P :=q_{1}^{t_{1}}\cdots q_{k+1}^{t_{k+1}} \, s_{1} \mbox{ and }
Q :=q_{1}^{t_{1}}\cdots q_{k+1}^{t_{k+1}} \, s_{2},
\]
for some $s_{1},s_{2}\in \Z$ to get
\begin{multline*}
3 \, q_{1}^{r_{1}} \cdots q_{k+1}^{r_{k+1}}=P^2+Q^2-PQ = \\
q_{1}^{2t_{1}}\cdots q_{k+1}^{2t_{k+1}}N_{\rho}(s_{1}+s_{2}\rho).
\end{multline*}
If $r_{i}=2t_{i}$ for all $i\in\{1,\ldots,k+1\}$, then $N_{\rho}(s_{1}+s_{2}\rho)=3$ and, by Lemma \ref{normassoconj}, $s_{1}+s_{2}\rho\sim \beta$. Hence, \begin{align*}
q_{1}^{t_{1}}\cdots q_{k+1}^{t_{k+1}}(s_{1}+s_{2}\rho) &\sim q_{1}^{t_{1}}\cdots q_{k+1}^{t_{k+1}} \, \beta,\\
P+Q \, \rho & \sim \beta \, \psi_{1}^{t_{1}} \, \overbar{\psi_{1}^{t_{1}}}\cdots \psi_{k+1}^{t_{k+1}} \, \overbar{\psi_{k+1}^{t_{k+1}}},\\
\beta\psi_{1}^{r_{1}}\cdots \psi_{k+1}^{r_{k+1}}&\sim \beta\psi_{1}^{t_{1}}\overbar{\psi_{1}^{t_{1}}}\cdots \psi_{k+1}^{t_{k+1}}\overbar{\psi_{k+1}^{t_{k+1}}}\mbox{, and}\\
\psi_{1}^{t_{1}}\cdots \psi_{k+1}^{t_{k+1}}&\sim \overbar{\psi_{1}^{t_{1}}}\cdots\overbar{\psi_{k+1}^{t_{k+1}}},
\end{align*}
which is impossible since $\psi_{i}\not\sim \overbar{\psi_{i}}$. 

We, again, partition the largest nonempty subset $I$ of $J=\{1,\ldots,k+1\}$ such that $r_{i}\neq 2t_{i}$ for any $i \in I$ into $I_{1}=\{i\in I \, : \,r_{i}>2t_{i}\}$ and $I_{2}=\{i \in I\,:\,r_{i}<2t_{i}\}$ to write
\[
3\prod_{i\in I_{1}}q_{i}^{r_{i}-2t_{i}} = \prod_{j\in I_{2}}q_{j}^{2t_{j}-r_{j}}N_{\rho}(s+t\rho).
\]

If $I_{2}$ is nonempty, then 
\[\frac{3\prod_{i\in I_{1}}q_{i}^{r_{i}-2t_{i}}}{\prod_{j\in I_{2}}q_{j}^{2t_{j}-r_{j}}}=N_{\rho}(s+t\rho),\]
which contradicts $N_{\rho}(s_{1}+s_{2}\rho)\in\Z$ since $3$ and any of the $q_{j}$s are distinct primes.

If $I_{2}$ is empty, then
\[
3\prod_{i\in I_{1}}q_{i}^{r_{i}-2t_{i}}= N_{\rho}(s_{1}+s_{2}\rho).
\]
By Lemma \ref{normassosiate}, 
\[
s_{1}+s_{2}\rho \sim \beta \, \prod_{i\in I_{1}}\psi_{i}^{r_{i}-2t_{i}} \mbox{ or } \overbar{\beta \, \prod_{i\in I_{1}}\psi_{i}^{r_{i}-2t_{i}}}.
\]
If $s_{1}+s_{2}\rho\sim \beta\,\prod_{i\in I_{1}}\psi_{i}^{r_{i}-2t_{i}}$, then 
\begin{align*}
q_{1}^{t_{1}}\cdots q_{k+1}^{t_{k+1}}(s_{1}+s_{2}\rho) &\sim q_{1}^{t_{1}}\cdots q_{k+1}^{t_{k+1}}\beta\prod_{i\in I_{1}}\psi_{i}^{r_{i}-2t_{i}},\\
 P+Q\rho &\sim  \beta \prod_{i\in J}\psi_{i}^{t_{i}}\overbar{\psi_{i}^{t_{i}}} \prod_{i\in I_{1}}\psi_{i}^{r_{i}-2t_{i}},\\
\beta\prod_{i\in J}\psi_{i}^{r_{i}} & \sim \beta \prod_{i\in I_{1}}\psi_{i}^{r_{i}-t_{i}}\prod_{i\in J}\overbar{\psi_{i}^{t_{i}}}\prod_{i\in J\setminus I_{1}}\psi_{i}^{t_{i}},\\
 \prod_{i\in J}\psi_{i}^{r_{i}} &\sim \prod_{i\in I_{1}}\psi_{i}^{r_{i}-t_{i}}\prod_{i\in J}\overbar{\psi_{i}^{t_{i}}}\prod_{i\in J\setminus I_{1}}\psi_{i}^{t_{i}},\\
 \prod_{i\in I_{1}}\psi_{i}^{t_{i}}\prod_{i\in J\setminus I_{1}}\psi_{i}^{r_{i}}&\sim \prod_{i\in J\setminus I_{1}}\psi_{i}^{t_{i}}\prod_{i\in J}\overbar{\psi_{i}^{t_{i}}},\\
 \prod_{i\in I_{1}}\psi_{i}^{t_{i}}\prod_{i\in J\setminus I_{1}}\psi_{i}^{r_{i}-t_{i}}& \sim \prod_{i\in J}\overbar{\psi_{i}^{t_{i}}},\\
   \prod_{i\in I_{1}}\psi_{i}^{t_{i}}\prod_{i\in J\setminus I_{1}}\psi_{i}^{t_{i}}& \sim \prod_{i\in J}\overbar{\psi_{i}^{t_{i}}},\\
    \prod_{i\in J}\psi_{i}^{t_{i}}& \sim \prod_{i\in J}\overbar{\psi_{i}^{t_{i}}},
\end{align*}
which is impossible since $\psi_{i}\not\sim \overbar{\psi_{i}}$.

If  $s_{1}+s_{2}\rho\sim \overbar{\beta\prod_{i\in I_{1}}\psi_{i}^{r_{i}-2t_{i}}}$, then 
\[
\prod_{i\in I_{1}}\psi_{i}^{r_{i}-t_{i}}\prod_{i\in J\setminus I_{1}}\psi_{i}^{t_{i}} \sim 
\prod_{i\in I_{1}}\overbar{\psi_{i}^{r_{i}-t_{i}}}\prod_{i\in J\setminus I_{1}}\overbar{\psi_{i}^{t_{i}}},\]
which is also impossible.

Let $r=3 \, q_{1}^{t_{1}}\cdots q_{k+1}^{t_{k+1}}$. We write 
\[
P := 3 \, q_{1}^{t_{1}}\cdots q_{k+1}^{t_{k+1}}s_{1} \mbox{ and } 
Q := 3 \, q_{1}^{t_{1}}\cdots q_{k+1}^{t_{k+1}}s_{2},
\]
for some $s_{1},s_{2}\in \Z$, and observe that 
\begin{align*}
3 \, q_{1}^{r_{1}}\cdots q_{k+1}^{r_{k+1}}&=P^2+Q^2-PQ\\
&=9 \, q_{1}^{2t_{1}}\cdots q_{k+1}^{2t_{k+1}}N_{\rho}(s_{1}+s_{2}\rho).
\end{align*}
Hence, in modulo $3$,
\[
q_{1}^{r_{1}}\cdots q_{k+1}^{r_{k+1}}= 3 \, q_{1}^{2t_{1}}\cdots q_{k+1}^{2t_{k+1}}N_{\rho}(s_{1}+s_{2}\rho)\equiv 0. 
\]
This is absurd since $q_{i}\equiv 1\Mod 3$ for all $i \in \{1,\ldots, k+1\}$ clearly means $q_{1}^{r_{1}}\cdots  q_{k+1}^{r_{k+1}}\equiv 1\Mod 3$. All cases considered, we conclude that $r:=\gcd(P,Q)=1$.

\section{Proof of Theorem \ref{jhjpsiku}}

Let $r\in\{0,1\}$ and let $r_{i}$ and $m$ be nonnegative integers. Let $\eta \sim\beta^{r}\psi_{1}^{r_{1}}\cdots \psi_{m}^{r_{m}}$. By Theorems \ref{psikua} and \ref{betapsikua}, we have $\gcd(a,b)=1$, making $a+b\rho$ a primitive Eisenstein integer. 

Conversely, let $\eta :=a+b\rho$, with $\gcd(a,b)=1$, whose prime decomposition is 
\begin{equation*}
\eta=(1+\rho)^{l}\beta^{r} \, \psi_{1}^{r_{1}} \, \cdots \, \psi_{m}^{r_{m}} \, \overbar{\psi_{1}}^{r'_{1}} \, \cdots \, \overbar{\psi}_{n}^{r'_{n}} \, p_{1}^{t_{1}} \, \cdots \, p_{k}^{t_{k}}.
\end{equation*}
Let $\eta$ be divisible by $\beta^{r}$ for $r \geq 2$ or divisible by prime integers $p^{t}$, where $p\equiv 2\Mod 3$ and $t\in \N$. If $\beta^{r}$, for $r\geq 2$, divides $\eta$, then $\eta = \beta^{r} \, \theta$ for some $\theta\in \Z[\rho]$. Since $\beta^{2s}\sim 3^s$, with $s\in \N$, then $\eta = \beta^{r} \, \theta \sim3^{s} \, \theta$ for $r=2s$ and $\eta=\beta^{r} \, \theta \sim 3^{s} \, \beta \, \theta$ for $r=2s+1$. Thus, $\gcd(a,b)\geq 3^{s}> 1$, which is absurd since $\gcd(a,b)=1$.

If $\eta$ is divisible by prime integers $p^{t}$, with $p\equiv 2\Mod 3$ and $t\in \N$, then $\eta= p^{t} \, \theta$ for some $\theta\in\Z[\rho]$. This implies $\gcd(a,b) \geq p^{t}>1$, which is a contradiction. It is impossible that $\eta$ is of the form \begin{equation*}
(1+\rho)^{l}\beta^{r} \, \psi_{1}^{r_{1}} \, \cdots \, \psi_{m}^{r_{m}} \, \overbar{\psi_{1}}^{r'_{1}} \, \cdots \,  \overbar{\psi}_{n}^{r'_{n}},
\end{equation*}
with $r\in\{0,1\}$ and $r_{i},r'_{j},m,n,l$ being nonnegative integers, since $q_{i}= \psi_{i} \, \overbar{\psi_{i}}$ implies $\gcd(a,b) \geq q_{i}> 1$. Thus, $\eta$ must be of the form $(1+\rho)^{l}\beta^{r} \, \psi_{1}^{r_{1}} \, \cdots \, \psi_{m}^{r_{m}}$, with $r\in\{0,1\}$ and $r_{i},l$ and $m$ being nonnegative integers.

\section{Proof of Theorem \ref{relnormhex}}
 
Let $\alpha=a+b\rho$ and $\theta=c+d\rho\in \Z[\rho]$.

To verify (i), we note that, if $a\leq 0\leq b$ or $b\leq 0\leq a$, then 
\begin{align*}
-ab&\geq 0,\\
a^2+b^2-ab&\leq a^2+b^2-2ab,\\
N_{\rho}(\alpha)&\leq (a-b)^2=(b-a)^2=(\wt_{\rm Hex}(\alpha))^2 \mbox{, and}\\
\sqrt{N_{\rho}(\alpha)}&\leq \wt_{\rm Hex}(\alpha).\
\end{align*}
If $0\leq a\leq b$ or $b\leq a\leq 0$, then 
\begin{align*}
a^2-ab&\leq 0,\\
N_{\rho}(\alpha)=a^2+b^2-ab&\leq b^2=(-b)^2=(\wt_{\rm Hex}(\alpha))^2,\\
\mbox{and } \sqrt{N_{\rho}(\alpha)}&\leq \wt_{\rm Hex}(\alpha).
\end{align*}
If $0\leq b\leq a$ or $a\leq b\leq 0$, then 
\begin{align*}
b^2-ab&\leq 0, \\
N_{\rho}(\alpha)=a^2+b^2-ab&\leq a^2=(-a)^2=(\wt_{\rm Hex}(\alpha))^2,\\
\mbox{and }\sqrt{N_{\rho}(\alpha)}&\leq \wt_{\rm Hex}(\alpha).
\end{align*}
Thus, $\sqrt{N_{\rho}(\alpha)}\leq \wt_{\rm Hex}(\alpha)$ for all $\alpha\in\Z[\rho]$. Besides that, if $a\leq 0\leq b$, then 
$\wt_{\rm Hex}(\alpha)=b-a\leq (b-a)b+a^{2}=N_{\rho}(\alpha)$.
If $0\leq a\leq b$, then 
$\wt_{\rm Hex}(\alpha)=b\leq a^2+b(b-a)=N_{\rho}(\alpha)$.
If $ a\leq b\leq 0$, then 
$\wt_{\rm Hex}(\alpha)=-a\leq -a(b-a)+b^2=N_{\rho}(\alpha)$. The other cases can be similarly confirmed to conclude that $\wt_{\rm Hex}(\alpha)\leq N_{\rho}(\alpha)$ for all $\alpha\in\Z[\rho]$.

To verify (ii) and (vii) we note that 
\begin{align*}
    \wt_{\rm Hex}(\alpha)&=\begin{cases}
     b-a,&\text{if}\,a\leq 0\leq b,\\
      a-b,&\text{if}\,b\leq 0\leq a,\\
     b,&\text{if}\,0\leq a\leq b,\\
     -b,&\text{if}\,b\leq a\leq 0,\\
     a,&\text{if}\,0\leq b\leq a,\\
   -a,&\text{if}\,a\leq b\leq 0,\\
 \end{cases} \mbox{ and }\\
 \wt_{\rm Hex}(\alpha\theta)&=\begin{cases}
     ad+bc-ac,&\text{if}\,ac\leq bd\leq ad+bc,\\
      ac-ad-bc,&\text{if}\,ad+bc\leq bd\leq ac,\\
     ad+bc-bd,&\text{if}\,bd\leq ac\leq ad+bc,\\
     bd-ad-bc,&\text{if}\,ad+bc\leq ac\leq bd,\\
     ac-bd,&\text{if}\,bd\leq ad+bc\leq ac,\\
   bd-ac,&\text{if}\,ac\leq ad+bc\leq bd.\\
 \end{cases}
\end{align*} 
For brevity, we give treatment for when $a\leq b$ and $c\leq d$, divided into nine cases. A similar treatment works for when $b\leq a$ and $d\leq c$, $a\leq b$ and $d\leq c$ as well as $b\leq a$ and $c\leq d$.
\begin{enumerate}
\item If $0\leq a\leq b$ and $0\leq c\leq d$, then 
\[
(b-a)(d-c)=bd+ac-ad-bc>0,
\]
making $ad+bc\leq ac+bd$. Since $ac\leq ad\leq ad+bc$ and $ac\leq bd$, we have $ac\leq bd\leq ad+bc$ or $ac\leq ad+bc\leq bd$. If $ac\leq bd\leq ad+bc$, then 
\[
b\leq bc\leq bc+a(d-c)\leq bd.
\]
If $ac\leq ad+bc\leq bd$, then 
\[
b\leq b(d-c)\leq bd-ac\leq bd. 
\]
\item If $0\leq a\leq b$ and $c\leq 0\leq d$, then $ac\leq 0\leq bd$ and $bc\leq 0\leq bd$. Since $ad+bc\leq ad\leq bd$, we have $ac\leq ad+bc\leq bd$ or $ad+bc\leq ac\leq bd$. If $ac\leq ad+bc \leq bd$, then 
\[
b\leq bd\leq bd-ac\leq bd-bc= b(d-c).
\]
If $ad+bc\leq ac\leq bd$, then 
\[
b\leq -bc\leq d(b-a)-bc=b(d-c)-ad\leq b(d-c).
\]
\item If $0\leq a\leq b$ and $c\leq d\leq 0$, then $ac,bd\leq 0$ and  $c(b-a)\leq d(b-a)\leq 0$, impying $ad+bc\leq ac+bd\leq ac$. Hence, $ad+bc \leq ac\leq bd$ or $ad+bc \leq bd\leq ac$. If $ ad+bc \leq ac\leq bd$, then 
\[
b\leq b(d-c)\leq bd-ad-bc=d(b-a)-bc\leq b(-c).
\]
If $ad+bc \leq bd\leq ac$, then 
\[
b\leq -bd\leq a(c-d)-bc\leq b(-c).
\]
\item If $a\leq 0\leq b$ and $0\leq c\leq d$, then 
\[
(b-a)(d-c)=ac+bd-ad-bc>0 \mbox{ and } ac\leq 0,
\]
implying $ad+bc\leq ac+bd\leq bd$. Since $ac\leq 0\leq bd$, we get $ac\leq ad+bc\leq bd$ or $ad+bc \leq ac\leq bd$. If $ ac\leq ad+bc\leq bd$, then 
\[
   b-a\leq (b-a)c=bc-ac\leq bd-ac\leq bd-ad=(b-a)d.
\]
If $ad+bc \leq ac\leq bd$, then 
\[
b-a\leq b(d-c)-ad= (b-a)d-bc \leq (b-a)d.
\]

\item If $a\leq 0\leq b$ and $c\leq 0\leq d$, then 
\[
ad+bc\leq 0\leq ac\leq bd \mbox{ or } 
ad+bc\leq 0\leq bd\leq ac.
\]
If $ ad+bc\leq 0\leq ac\leq bd$, then 
\[
   b-a\leq (b-a)d\leq (b-a)d-bc\leq (b-a)(d-c).
\]
If $ad+bc \leq 0\leq bd\leq ac$, then 
\[
b-a\leq (b-a)(-c)\leq ac-ad-bc\leq (b-a)(d-c).
\]

\item If $a\leq 0\leq b$ and $c\leq d\leq 0$, then $bd\leq 0\leq ac$ and $(b-a)(d-c)=bd+ac-ad-bc\geq 0$. Hence, $ad+bc \leq ac+bd\leq ac$, which means $ ad+bc \leq bd\leq ac$ or $bd \leq ad+bc\leq ac$. If $ad+bc\leq bd\leq ac$, then 
\[
   b-a\leq b(-c)-a(d-c)=-c(b-a)-ad\leq -c(b-a).
\]
If $bd \leq ad+bc\leq ac$, then 
\[
b-a\leq b(-d)-a(-c)=ac-bd\leq ac-bc=-c(b-a).
\]

\item If $a\leq b\leq 0$ and $0\leq c\leq d$, then 
\[(b-a)(d-c)=bd+ac-ad-bc\geq 0,\] which implies that $ad+bc\leq bd+ac$, meaning either $ad+bc\leq ac\leq bd\leq 0$ or $ad+bc\leq bd\leq ac\leq 0$. If the former, then
\[
   -a\leq  -ac\leq bd-ad-bc=b(d-c)-ad\leq -ad.
\]
If the latter, then 
\[
-a\leq -a(d-c)\leq a(c-d)-bc=c(a-b)-ad\leq -ad .
\]
\item If $a\leq b\leq 0$ and $c\leq 0\leq d$, then 
\[
(b-a)(d-c)=bd+ac-ac-bc\leq 0,
\]
implying that $ad+bc\leq ac+bd\leq ac$. If $bd\leq ad+bc \leq ac$, then 
\[
   -a\leq (-a)(-c)\leq ac-bd\leq ac-ad\leq -a(d-c).
\]
If $ad+bc \leq bd\leq ac$, then 
\[
-a\leq -ad\leq c(a-b)-ad\leq ac-ad=-a(d-c).
\]
\item If $a\leq b\leq 0$ and $c\leq d\leq 0$, then $(d-c)(b-a)\geq 0$, making $0\leq ad+bc\leq ac+bd$. Hence, $0\leq bd\leq ad+bc \leq ac$ or $0\leq bd\leq ac\leq ad+bc $. If the former, then 
\[
   -a\leq -a(d-c)=ac-ad\leq ac-bd\leq (-a)(-c).
\]
If the latter, then 
\[
-a\leq (-a)(-d)=ad\leq ad+b(c-d)\leq (-a)(-c)
\]
\end{enumerate}

The respective justifications for (iii) and (iv) are clear by the definition of hexagonal weight.

For (v), let $\alpha\sim n+ k \, \rho$ and $\theta\sim n+ \ell \, \rho$ for some $k, \ell \in \{0,1,\ldots,n-1\}$. By definition, $\wt_{\rm Hex}(\alpha)=\wt_{\rm Hex}(\theta)=n$. Conversely, letting $\wt_{\rm Hex}(\alpha)=\wt_{\rm Hex}(\theta)=n$ yields $\alpha\sim n+k \, \rho$ and $\theta\sim n+ \ell \, \rho$ for some $k,\ell \in \{0,1,\ldots,n-1\}$.

To confirm (vi), let $N_{\rho}(\alpha)=N_{\rho}(\theta)$. By Theorem \ref{normassosiate}, we have $\alpha\sim \theta$ or $\alpha\sim\overbar{\theta}$. Hence, $\alpha\sim n+k \, \rho$ and $\theta\sim n+ \ell \, \rho$ for some $k,\ell \in \{0,1,\ldots,n-1\}$. By (v), $\wt_{\rm Hex}(\alpha)=\wt_{\rm Hex}(\theta)$.

\section*{Acknowledgments}
A. Hadi is supported by the Indonesian Center for Higher Education Funding (BPPT) and by the Endowment Funds for Education, known by its abbreviation LPDP in Bahasa Indonesia on decree no. 00082/J5.2.3./BPI.06/9/2022. 


\end{document}